\documentclass[twocolumn]{aastex631}
\usepackage{amsmath}
\usepackage{CJK}
\usepackage{xcolor}
\usepackage{graphicx}
\usepackage{hyperref, array}
\usepackage{booktabs} 

\usepackage{CJK}

\usepackage{bm}
\usepackage{soul}

\defcitealias{goldreich_toward_1995}{GS95}
\defcitealias{boldyrev_2006}{B06}
\defcitealias{chandran_intermittency_2015}{CSM15}
\defcitealias{Podesta_2009}{P09}

\newcommand\alfvenic{Alfv\'enic}
\newcommand\elsasser{Els\"asser}

\begin{document}
\begin{CJK*}{UTF8}{gbsn}

\title{Scale-Dependent Dynamic Alignment in MHD Turbulence: Insights into Intermittency, Compressibility, and Imbalance Effects}

\author[0000-0002-1128-9685]{Nikos Sioulas }
\affiliation{Department of Earth, Planetary, and Space Sciences, University of California, Los Angeles, CA, USA }

\author[0000-0002-2381-3106]{Marco Velli}
\affiliation{Department of Earth, Planetary, and Space Sciences, University of California, Los Angeles, CA, USA}

\author[0000-0003-4177-3328]{Alfred Mallet}
\affiliation{Space Sciences Laboratory, University of California, Berkeley, CA 94720-7450, USA}

\author[0000-0002-4625-3332]{Trevor A. Bowen}
\affiliation{Space Sciences Laboratory, University of California, Berkeley, CA 94720-7450, USA}

\author[0000-0003-4177-3328]{B. D. G. Chandran}
\affiliation{Space Science Center and Department of Physics, University of New Hampshire, Durham, NH 03824, USA}

\author[0000-0002-2582-7085]{Chen Shi (时辰)}
\affiliation{Department of Earth, Planetary, and Space Sciences, University of California, Los Angeles, CA, USA}

\author[0000-0003-0562-6574]{S. S. Cerri}
\affiliation{Universit\'e Côte d’Azur, Observatoire de la Côte d’Azur, CNRS, Laboratoire Lagrange, Bd de l’Observatoire, CS 34229, 06304 Nice5cedex 4, France}

\author[0000-0002-8921-3760]{Ioannis Liodis}
\affiliation{Department of Physics,  Aristotle University of Thessaloniki\\
GR-52124 Thessaloniki, Greece }

\author[0000-0002-8475-8606]{Tamar Ervin}
\affiliation{Department of Physics, University of California, Berkeley, Berkeley, CA 94720-7300, USA}
\affiliation{Space Sciences Laboratory, University of California, Berkeley, CA 94720-7450, USA}

\author[0000-0001-5030-6030]{Davin E. Larson}
\affiliation{Space Sciences Laboratory, University of California, Berkeley, CA 94720-7450, USA}

\begin{abstract}

Scale-Dependent Dynamic Alignment (SDDA) in the polarization of Els\"asser field fluctuations is theorized to suppress nonlinearities and modulate the energy spectrum. Despite its significance, empirical evidence for SDDA within the inertial range of solar wind turbulence has been conspicuously absent. We analyze a large dataset of homogeneous intervals from the WIND mission to assess the effects of compressibility, intermittency, and imbalance on SDDA. Our key findings are as follows: SDDA is consistently evident at energy-containing scales, but a trend towards misalignment is, on average, observed at inertial scales. 
While compressible fluctuations do not exhibit any increasing alignment, their impact on the average behavior of SDDA is negligible. Therefore their mixing with incompressible fluctuations cannot explain the trend of misalignment at inertial scales.
The alignment angles exhibit an inverse correlation with the intensity of field gradients. This may stem from ``anomalous'' and/or ``counterpropagating'' wave packet interactions. Regardless, this observation indicates that the physical origin of alignment arises from the mutual shearing of the {\elsasser} fields during imbalanced ($\delta \boldsymbol{z}^{\pm} \gg \delta \boldsymbol{z}^{\mp}$) wave-packet interactions. Stringent thresholding on field gradient intensity reveals SDDA signatures within a significant portion of the inertial range. The scaling of the alignment angle in the Els\"asser increments,  $\Theta^{z}$, becomes steeper with increasing global {\alfvenic} imbalance, while the angle between magnetic/velocity field increments. $\Theta^{ub}$, becomes shallower. Only $\Theta^{ub}$ is correlated with global Els\"asser imbalance, becoming steeper as the imbalance increases. Additionally, increasing alignment in $\Theta^{ub}$ extends deep into the inertial range of balanced intervals but breaks down at large scales for imbalanced ones. We present a simplified theoretical analysis and model the effects of high-frequency, low-amplitude noise in the velocity field on SDDA measurements. Our analysis indicates that noise can significantly affect alignment angle measurements even at very low frequencies, with the impact growing as global imbalance increases.

\keywords{Solar wind; Interplanetary turbulence; Magnetohydrodynamics; Space plasmas}

\end{abstract}

\section{Introduction}\label{sec:intro}

\par In the framework of reduced magnetohydrodynamics (RMHD), applied to
highly conducting, magnetized plasmas threaded by a large-scale background magnetic field, $\boldsymbol{B}_{0}$, on scales larger than the ion gyroradius, $\rho_i$, the  \cite{Alfven_1942} polarized fluctuations decouple from the compressive cascade and and obey the equation \citep{Kadomtsev_Pogutse, schekochihin_astrophysical_2009}

\begin{equation}
\partial_{t}\boldsymbol{z}_{\perp}^{\mp} \pm V_{A} \partial_{z} \boldsymbol{z}_{\perp}^{\mp} + \boldsymbol{z}_{\perp}^{\pm} \cdot \nabla_{\perp} \boldsymbol{z}_{\perp}^{\mp} = - \nabla_{\perp}p,
\label{fin_ch_eq:1}
\end{equation} 
where $\boldsymbol{V}_{a} = \boldsymbol{B}_{0}/ \sqrt{4 \pi \rho}$  is the Alfv\'en speed, $\boldsymbol{z}_{\perp}^{\pm} = \boldsymbol{u}_{\perp} \mp\boldsymbol{b}_{\perp}/ \sqrt{4 \pi \rho}$ are the \cite{Elsasser_1950} variables, $\boldsymbol{u}_{\perp}$ and $\boldsymbol{b}_{\perp}$ represent transverse (to $\boldsymbol{B}_{0}$), small-amplitude ($|\boldsymbol{b}_{\perp}| /|\boldsymbol{B}_{0}| \ll 1$) velocity and magnetic-field perturbations, respectively, and the total pressure, $p$, can be determined by taking the (perpendicular) divergence of Equation~(\ref{fin_ch_eq:1}) and imposing the condition $\nabla_{\perp} \cdot \boldsymbol{z}_{\perp}^{\pm} = 0$. In this framework, nonlinearity arises from the collisions of counterpropagating wave packets, resulting in their distortion and fragmentation \citep{iroshnikov_turbulence_1963, kraichnan_inertial-range_1965}.  The effectiveness of the nonlinear interactions hinges on the relative importance of the linear and nonlinear terms in Equation~\ref{fin_ch_eq:1}, which can be quantified by the nonlinearity parameter, $\chi^{\pm} =  (\ell^{\pm}_{||, \lambda}/ \lambda)/(\delta z_{\lambda}^{\pm}/V_{a})$ which compares the linear wave propagation time $\tau^{\pm}_{A} = \ell^{\pm}_{||}/ V_{A}$ and non-linear decorrelation time $\tau^{\mp}_{nl} \sim \lambda/\delta \boldsymbol{z}^{\pm}_{\lambda}$, where, ($\ell_{||} \sim 1/ k_{||}$),  ($\lambda \sim 1/ k_{\perp}$)  the coherence lengths parallel and perpendicular to $\boldsymbol{B}_{0}$.

The interaction of Alfv\'enic wavepackets, described by the dispersion relation $\omega = k_{||}v_{A}$, where $k_{||}$ is the wavevector parallel to $B_0$, gives rise to a distinctive feature of MHD turbulence wherein the flux of energy is primarily directed towards smaller scales perpendicular to the magnetic field
. The cascade is strongly anisotropic, and even more so at smaller scales, resulting in structures satisfying $\ell_{||} \gg \lambda$ \citep{Robinson_1971, shebalin_matthaeus_montgomery_1983, higdon_anisotropic, 1994_oughton_matthaeus_priest}. The anisotropy amplifies the non-linearity of the interactions \citep{2000_Gatlier}. Consequently, even in the case where the system was forced at large scales such that $\chi \ll 1$ (referred to as the ``weak turbulence'' regime), it inevitably transitions, at sufficiently small scales, to a state where within a single collision, the wave packet undergoes deformation of a magnitude comparable to its own, $\chi \sim 1$, i.e.,  the cascade time
$\tau_{c}$ and dynamical $\tau_{d} = 1/wk_{\perp}$ timescales are similar. Since the perturbation frequency $\omega$ has a lower bound due to an uncertainty relation $\omega \tau_{c}  >1$, the cascade is forced to remain in the $\chi \sim 1$ regime. This realization forms the basis of what is formally known as the ``critical balance'' (CB) conjecture \citep{goldreich_toward_1995}, denoted hereafter as \citetalias{goldreich_toward_1995}. In the framework of \emph{balanced} turbulence, i.e., assuming equal energy fluxes, $\epsilon^{\pm}$, in counterpropagating wavepackets, CB implies an anisotropic scaling relationship between the parallel and perpendicular wavevectors, 
specifically $k{||} \propto k_{\perp}^{2/3}$, leading to expected field-perpendicular and field-parallel energy
spectra of the form $E(k_{\perp}) \propto k_{\perp}^{-5/3}$ and $E(k_{||}) \propto k_{||}^{-2}$, resepectively.

While in situ observations of the solar wind appeared consistent with the phenomenology proposed by \citetalias{goldreich_toward_1995} \citep{horbury_anisotropic_2008, wicks_anisotropy1}, numerical simulations of homogeneous, incompressible MHD turbulence revealed a significant discrepancy. Specifically, the observed spectral index perpendicular to the locally defined background magnetic field, $\boldsymbol{B}_{\ell}$, was closer to $-3/2$ \citep{Maron_2001, 2003PhRvE..67f6302M, Muller_grappin_2005}. These findings prompted refinements to the \citetalias{goldreich_toward_1995} model \citep{Boldyrev_2005, 2005PhPl...12i2310G, 2006_Beresnyak, 2007PhPl...14b2304G}.

One approach to addressing this discrepancy was the incorporation of scale-dependent dynamic alignment (SDDA) into the \citetalias{goldreich_toward_1995} framework.  \cite{boldyrev_2006}, henceforth \citetalias{boldyrev_2006}, proposed that  a self-consistent mechanism that results in increasing alignment would modify the spectral slope of the field-perpendicular inertial range energy spectrum from the -5/3 Kolmogorov slope to the numerically observed -3/2 slope. Specifically, \cite{boldyrev_2006} suggests that the non-linearity in MHD turbulence is reduced due to the increasing alignment between $\delta\boldsymbol{v}_{\perp}$ and $\delta\boldsymbol{b}_{\perp}$ as $\lambda$ decreases, with $\theta^{ub}_{\lambda} \sim \delta b/v_A \propto \lambda^{1/4}$. At small scales, this leads to the emergence of three-dimensional anisotropic eddies characterized by $\ell_{||} \gg \xi \gg \lambda$, where $\xi$ represents the coherence length in the direction of $\delta \boldsymbol{b}$, and $\lambda$ is perpendicular to both $\boldsymbol{B}_{\ell}$ and $\delta \boldsymbol{b}$.

A substantial body of numerical studies on homogeneous MHD turbulence has provided evidence supporting the scale-dependence of certain alignment measures across a sizable portion of the inertial range \citep{masson_2006, Perez_2012, Perez_2014, 2015Mallet, chandran_intermittency_2015, 2022_cerri}. However, concerns have been raised suggesting that the observed alignment may be a finite-range effect intrinsically linked to dynamics occurring at the outer scale \citep{2012_Beresnyak}. For instance, \citet{2012_Beresnyak} interpret these signatures based on the idea that MHD turbulence is much less local in k-space compared to hydrodynamic turbulence \citep[see, e.g.,][]{Beresnyak_2011, Schekochihin_2022}. Consequently, the driving mechanism does not fully replicate the properties of the inertial range, and the transition to asymptotic statistics is broad, causing many quantities to appear scale-dependent as they adjust to the asymptotic regime. Moreover, the \citetalias{boldyrev_2006} model has faced criticism for violating the rescaling symmetry of the RMHD equations \citep{2012_Beresnyak}. As a result, an ongoing debate persists regarding whether this numerical evidence accurately reflects the scale-dependent dynamic alignment angle in the asymptotic state of the inertial range \citep{Mason_2011, 2012_Beresnyak, Perez_2014}.

The theory of SDDA has been revisited by \citet[][hereafter, \citetalias{chandran_intermittency_2015}]{chandran_intermittency_2015}, refining it in a manner that aligns SDDA with the rescaling symmetry of RMHD. Their model distinguishes between two archetypal types of alfv\'enic interactions: imbalanced ($\delta \boldsymbol{z}^{\pm} \gg \delta \boldsymbol{z}^{\mp}$) and balanced ($\delta \boldsymbol{z}^{\pm} \sim \delta \boldsymbol{z}^{\mp}$). They develop an approximate theory which demonstrates that in imbalanced interactions, the mutual shearing of the interacting wave packets leads to the rapid cascading of the subdominant field to smaller scales. This occurs as it rotates into alignment with the dominant field, notably without distorting their amplitudes.  Consequently, at any given scale, wave packets subjected to the smallest number of balanced collisions—those that alter both amplitude and coherence lengths—roughly retain their outer scale amplitudes, resembling three-dimensional anisotropic current sheet structures \citep[see also][]{Howes2015, Mallet_2017}. This indicates that as the cascade progresses toward smaller scales, the fluctuating energy becomes confined within an increasingly smaller volume fraction, thus establishing a link between SDDA and intermittency \citep{oboukhov_1962, kolmogorov_refinement_1962}.

The framework established by \citetalias{chandran_intermittency_2015} has garnered significant support from the studies conducted by \cite{2015Mallet, 2016_mallet}, who employed RMHD numerical simulations to demonstrate that the degree of alignment at any given scale increases with the fluctuation amplitude. This finding indicates that alignment angles exhibit intermittency rather than scale invariance \citep[see also][]{Beresnyak_2006}. Moreover, scaling predictions derived from these phenomenological models, particularly those related to higher-order moments and alignment angles, show strong agreement with numerical simulations of forced, homogeneous, and isotropic MHD turbulence \cite{chandran_intermittency_2015, 2016_mallet, Mallet_2017, 2023_Chen_compres, Dong_22_largest_mhd_turb}.

However, in-situ solar wind observations reveal certain deviations from the model predictions. For example, while inertial range scalings of the perpendicular components for highly Alfv\'enic intervals tend towards the predicted -3/2 value, for both magnetic field and velocity fluctuations, the former  steepen towards -5/3 in balanced streams \citep{2013_Chen_residual, Sioulas_2022A_preferential, mcintyre2023properties}. Moreover, emerging evidence indicates the presence of two sub-regimes within the traditionally defined inertial range, particularly prominent in strongly imbalanced streams and when considering anisotropic spectra in locally defined frames \citep{Wicks_2011, Sioulas_2022_intermittency, Wu2022OnTS, Sioulas_2023_anisotropic}. Although 3D anisotropic eddies are evident at the smaller scale end of the inertial range, anisotropy qualitatively changes across scales \citep{Chen_2012ApJ, Verdini_3D_2018, sioulas2024higherorder}.

Finally, in-situ observations of SDDA in the solar wind remain inconclusive. While SDDA is evident at large, energy-containing scales in the $\textit{1/f}$ range, a shift occurs in the inertial range ($\lambda \approx 10^{4}d_i$), where increasing misalignment of fluctuations is observed towards smaller scales \cite{Podesta_2009, Hnat_scale_free_2011, wicks_alignmene_2013, 2018_Parashar, Parashar_2020}. The increasing misalignment at inertial scales persists even in intervals selected to minimize solar wind expansion effects \cite{Verdini_3D_2018}, although in this case, the 3D anisotropic inertial range spectral scaling and topology of eddies align with proposed models \cite{boldyrev_2006, Chandran_2015, Mallet_2017}.

These observations could suggest that additional physical mechanisms beyond standard homogeneous MHD are necessary to fully explain the observed properties of solar wind turbulence. Such mechanisms include, but are not limited to, the breaking of local anisotropy due to the additional radial symmetry axis introduced by expansion \citep{Dong_2014, 2015_verdini, Verdini_3D_2018}, the influence of non-Alfv\'enic interactions in the turbulent cascade \citep[see, e.g.,][]{Chapman_2007, bowen2021nonlinear}, the spherically polarized nature of Alfv\'enic fluctuations \citep{Matteini_2014, Mallet_2021, Matteini_2024}, and the imbalance in the fluxes of counterpropagating wave packets \citep{Lithwick_2007_imbalanced_critical_balance, Chandran_2008_cross_hel, Perez_2009, Schekochihin_2022}.

In recent years, considerable efforts have been made to gauge the impact of imbalance, compressibility, intermittency, and expansion on the statistical properties of MHD\ turbulence \citep{Lithwick_2001, Hnat_compress, Salem_intermittency, Lithwick_2007_imbalanced_critical_balance, Chandran_2008_cross_hel, Podesta_2010, Podesta_Bhattacharjee_2010, Gang_li_2011, 2013_Chen, wicks_alignmene_2013, Matteini_2014, 2018_Bowen, 2018_Bowen_residual, 2019ApJ_Shoda, Vech_2016, meyrand_2021, Sioulas_2022_intermittency, sioulas_statistical_2022, 2023_Dunn, 2023_Chen_compres, WANG_2023}. However, previous investigations, particularly those focusing on in-situ observations, have not thoroughly explored these effects on the statistical properties of SDDA. This work aims to bridge this gap.

\begin{figure*}
     \centering
     \includegraphics[width=1\textwidth]{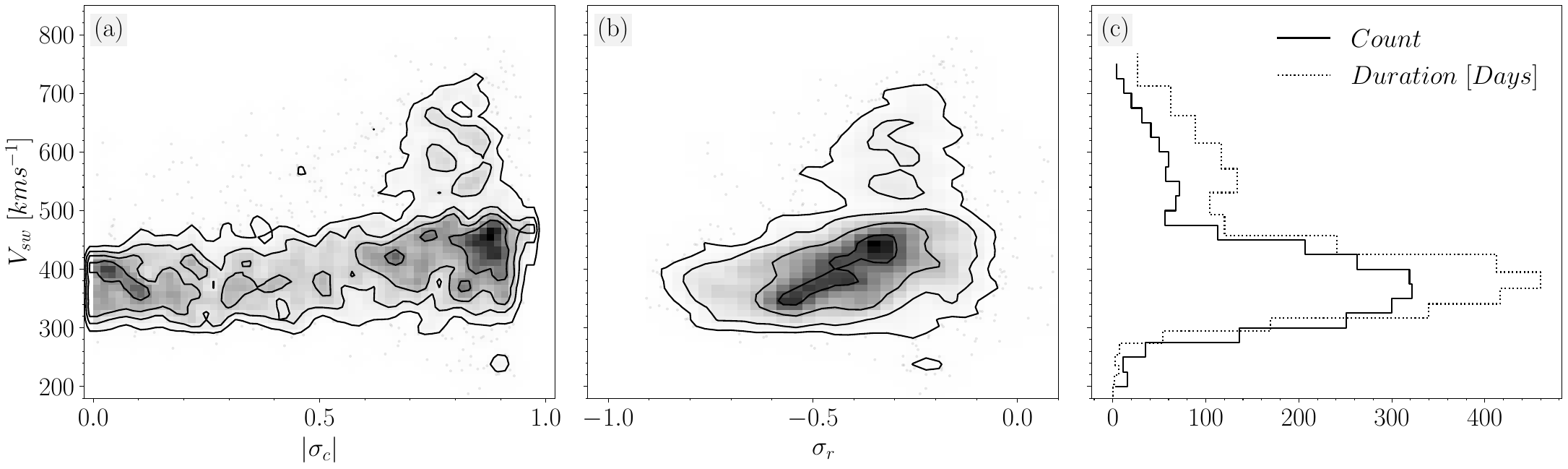}
     \caption{ Joint probability distribution of $V_{\text{sw}}$ and (a) $\sigma_c$, (b) $\sigma_r$ for the homogeneous intervals selected for our analysis. $\sigma_c$ and $\sigma_r$ are computed based on fluctuations at the scale of $2\cdot 10^{4} d_i$. In Panel (c), a histogram displays interval counts and cumulative durations relative to $V_{\text{sw}}$. }
     \label{fig:overview}
\end{figure*}

To this end, we conduct a statistical analysis of carefully selected solar wind intervals spanning a 28-year period from WIND observations. We utilize scale and time-dependent proxies to isolate the properties of interest, namely compressibility, intermittency, and imbalance, enabling us to quantify their effects on SDDA measurements.

\par The remainder of this paper is structured as follows: Section~\ref{sec:Data} details the selection and processing of the data. Section~\ref{sec:data_analysis} provides background on the methods applied for this analysis.  The results of this study are presented in Section~\ref{sec:Results}, followed by a discussion of these results and the conclusions in Section~\ref{sec:Conclusions}.

\section{Data Selection} \label{sec:Data}

We utilize data collected by instruments aboard the WIND spacecraft, positioned at Earth's L1 Lagrange point, approximately 1 AU from the Sun.

This study primarily analyzes magnetic field and ion measurements obtained from the MFI instrument \citep{1995_lepping} and the 3DP/PESA-L instrument \citep{1995SSRv...71..125L}, which provide data with resolutions of 0.1 seconds and 3 seconds, respectively. It should be noted that, while a 3-second dataset is available for the magnetic field timeseries, an issue has been identified with this specific dataset, as detailed in~\ref{appendix:CDWEB}. Consequently, the full-resolution MFI data were downsampled—following the application of a low-pass Butterworth filter to mitigate aliasing—to match the temporal resolution of the ion moment data.

In the preliminary phase of our analysis, we conducted a meticulous visual inspection of the WIND timeseries data spanning from January 1, 1995, to August 1, 2023. Based on this examination, we classified the data into three distinct categories:
\begin{enumerate}
\item Slow Alfv\'enic wind: $V_{sw} \leq 450$ km/s, $\sigma_c \geq 0.85$
\item Slow non-Alfv\'enic wind: $V_{sw} \leq 450$ km/s, $\sigma_c \leq 0.85$
\item Fast Alfv\'enic wind: $V_{sw} \geq 500$ km/s, $\sigma_c \geq 0.85$
\end{enumerate}

Throughout the visual inspection, strict adherence was maintained to several selection criteria:
\begin{enumerate}
\item A minimum interval duration of 6 hours was mandated,
\item The stability of plasma parameters ($V_{sw}$, $n_p$, $\beta$, $\Theta_{VB}$, $\sigma_c$, $\sigma_r$, $\text{sign}(B_x)$) was maintained, ensuring that these parameters did not undergo significant variations throughout the selected interval,
\item Intervals showing clear indications of transient events, like Coronal Mass Ejections or Heliospheric Current Sheet crossings, were omitted,
\item Intervals were restricted to those with a maximum of 2\% missing magnetic field data and 5\% missing plasma data.
\end{enumerate}

The selection process resulted in a total of 2335 intervals. The distribution of interval characteristics within our dataset is visualized in Fig.~\ref{fig:overview}. Panels (a) to (c) depict 2D histograms illustrating the solar wind speed in correlation with normalized cross helicity ($\sigma_c$), normalized residual energy ($\sigma_r$), both evaluated at a scale of $\ell \approx 2\cdot 10^{4} d_i$. Finally, column (c) provides a histogram portraying the number of selected intervals (solid line) and their cumulative duration associated with different $V_{sw}$ values.

Furthermore, to facilitate a more direct comparison with \citep{Podesta_2009} (hereinafter referred to as \citetalias{Podesta_2009}), we examined four intervals spanning: (a) January 1, 1995, to July 29, 1995, (b) May 15, 1996, to August 16, 1996, (c) January 8, 1997, to June 9, 1997, (d) August 23, 2000, to February 15, 2001. For an exhaustive discussion regarding the properties of the chosen intervals, readers are directed to \citetalias{Podesta_2009}. An in-depth comparison between our results and those of \citetalias{Podesta_2009} is provided in Appendix~\ref{appendix:1}.

\section{Data Analysis}\label{sec:data_analysis}

\subsection{Estimating 5-point increments} \label{subsec:5pt_increments}

Assuming the validity of \cite{taylor_spectrum_1938} hypothesis the 2-point increments for a field $\boldsymbol{\phi}$ at a specific spatial lag, where $\boldsymbol{\ell} = \tau \cdot \boldsymbol{V}_{sw}$ and $\boldsymbol{V}_{sw}$ represents the solar wind speed, can be estimated as:

\begin{equation}
\delta \boldsymbol{\phi} = \boldsymbol{\phi}(\boldsymbol{r} + \boldsymbol{\ell}) - \boldsymbol{\phi}(\boldsymbol{r}),
\label{fin_ch_eq:2}
\end{equation}

However, in the context of estimating five-point increments, Equation~~\ref{fin_ch_eq:2} necessitates a redefinition of $\delta \boldsymbol{\phi}$ to:

\begin{equation}
\begin{aligned}[b]
    \delta \boldsymbol{\phi} = ~& [\boldsymbol{\phi}(\boldsymbol{r}-2\boldsymbol{\ell})-4 \boldsymbol{\phi}(\boldsymbol{r}-\boldsymbol{\ell})+6 \boldsymbol{\phi}(\boldsymbol{r})\\ & \qquad -4\boldsymbol{\phi}(\boldsymbol{r}+\boldsymbol{\ell})+ \boldsymbol{\phi}(\boldsymbol{r}+2\boldsymbol{\ell})]/\sqrt{35}.
\end{aligned}
\label{fin_ch_eq:5}
\end{equation}

Using the five-point method, we estimate the scale-dependent increments of the magnetic field—in velocity units—as:

\begin{equation}
\delta \boldsymbol{b} = \frac{\delta \boldsymbol{B}}{\sqrt{ \mu_{0} \rho}},
\end{equation}

where $\rho = 1.16 m_p n_p$, $\mu_{0}$ is the permeability of free space, and $m_p$ and $n_p$ represent the proton mass and a 1-minute rolling average of the number density, respectively. The factor 1.16 accounts for alpha particles, assuming $n_a/n_p = 0.04$ \citep{PODESTA_2009_SDDA}. It is noteworthy that similar results were obtained even when this analysis was performed without the prefactor.

 The perpendicular components of the increments are defined by

\begin{equation}
\delta \boldsymbol{\phi}_{\perp} = \delta\boldsymbol{\phi} - (\delta \boldsymbol{\phi} \cdot \hat{\boldsymbol{z}} ) \hat{\boldsymbol{z}},
\label{fin_ch_eq:6}
\end{equation}
where, $\hat{\boldsymbol{z}} = \boldsymbol{B_{\ell}}/|\boldsymbol{B_{\ell}}|$, is a  unit vector in the direction of the localy-defined scale-dependent background magentic field estimated as

\begin{equation}
\begin{aligned}[b]
\boldsymbol{B}_{\ell} = ~& [\boldsymbol{B}(\boldsymbol{r} - 2 \boldsymbol{\ell}) + 4\boldsymbol{B}(\boldsymbol{r} - \boldsymbol{\ell}) + 6\boldsymbol{B}(\boldsymbol{r}) \\ & \qquad + 4\boldsymbol{B}(\boldsymbol{r} + \boldsymbol{\ell}) + \boldsymbol{B}(\boldsymbol{r} + 2\boldsymbol{\ell})]/16.
\label{fin_ch_eq:7a}
\end{aligned}
\end{equation}

For our analysis, we evaluated both the 2-point and 5-point increment methods. Across the range of scales considered, both methods yielded similar results. However, the 2-point method has been shown to produce erroneous results at scales where the turbulent cascade leads to steep scalings \citep{ sioulas2024higherorder}. Therefore, for consistency with our previous and forthcoming studies, we have chosen to proceed with the 5-point increments approach.

\begin{figure*}
     \centering
     \includegraphics[width=1\textwidth]{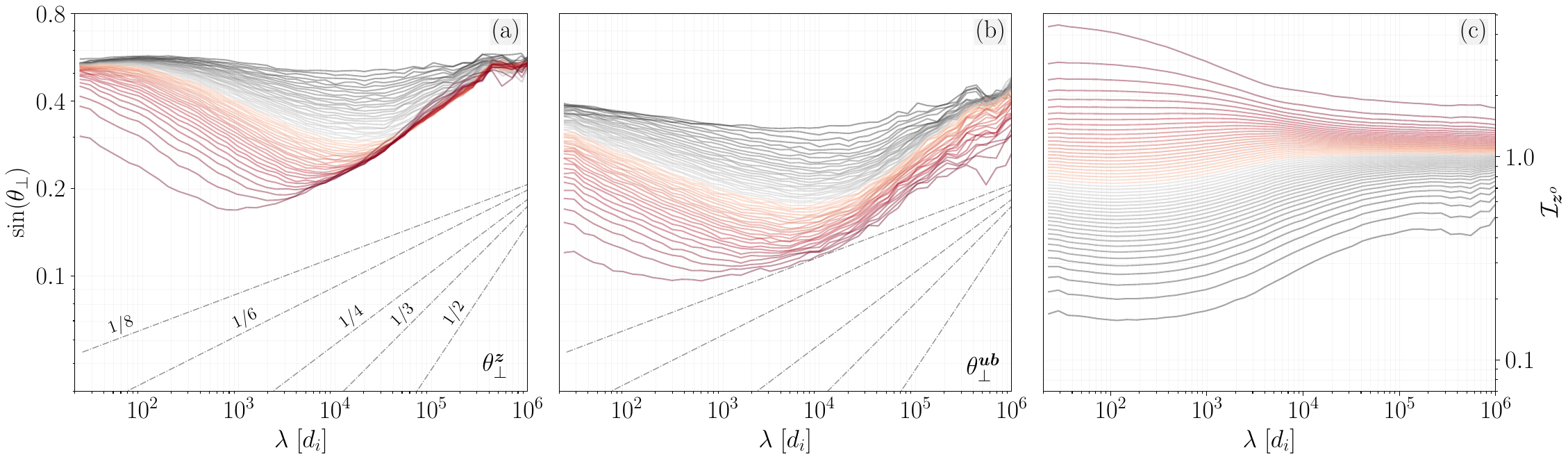}

     \caption{(a) $\sin( \theta_{\perp}^{\boldsymbol{z}})$, (b) $\sin(\theta_{\perp}^{\boldsymbol{ub}})$. Each subplot categorizes alignment angles into 50 bins based on $\mathcal{I}_{\boldsymbol{z^{o}}}$, with the bins ranging from the lowest PVI values (black) to the highest PVI values (red), as shown in panel (c). Reference lines depict the scaling parameter $\theta \propto \lambda^{\alpha}$, facilitating comparative analysis.}
     \label{fig:sdda_pvi}
\end{figure*}

\subsection{Alignment Angles} \label{subsec:align_angles}

Conventionally, the alignment angle between the perpendicular components of field increments, $\delta \boldsymbol{\phi}_{\perp}$ and $\delta \boldsymbol{\psi}_{\perp}$, is estimated using:

\begin{equation}\label{fin_ch_eq:7}
\sin(\theta^{\phi \psi}_{\perp}) = \left\langle \frac{|\delta \boldsymbol{\phi}_{\perp} \times \delta\boldsymbol{\psi}_{\perp}|}{|\delta \boldsymbol{\phi}_{\perp}| |\delta \boldsymbol{\psi}_{\perp}|} \right\rangle.
\end{equation}

This definition can be applied to estimate the angle between $\delta \boldsymbol{u}_{\perp}$ and $\delta \boldsymbol{b}_{\perp}$, denoted as $\theta_{\perp}^{ub}$, as well as the angle $\theta_{\perp}^{z}$, which is between the perpendicular components of the increments in the outwardly and inwardly propagating modes:

\begin{equation}
\delta \boldsymbol{z}_{\perp}^{o,i} = \delta\boldsymbol{v}_{\perp} \mp \text{sign}(\boldsymbol{B}^{r}_{\ell}  )\delta \boldsymbol{b}_{\perp},
\end{equation}\label{fin_ch_eq:8a}

where, $\boldsymbol{B}^{r}_{\ell}$ is  the radial component of $\boldsymbol{B}_{\ell}$, in RTN coordinates \citep{FRANZ2002217}. The local average of $B^{r}$ assists in determining the polarity of the radial magnetic field \citep{shi_alfvenic_2021}. It is worth noting that we consider data in the GSE coordinate system; therefore, we use $B^{r} = -B^{x}$.

An alternative definition of these angles is obtained by separately averaging the numerator and denominator:

\begin{equation}\label{fin_ch_eq:8}
\sin (\tilde{\theta}^{\phi \psi}_{\perp}) = \frac{\langle |\delta \boldsymbol{\phi}_{\perp} \times \delta\boldsymbol{ \psi}_{\perp}| \rangle}{ \langle |\delta \boldsymbol{\phi}_{\perp}| |\delta \boldsymbol{\psi}_{\perp}| \rangle}.
\end{equation}

This approach allows for the identification of ``dynamically relevant''  fluctuations, as the averaging procedure considers that, at a given $\lambda$, fluctuations with amplitudes near the RMS value are the primary contributors to turbulent dynamics \citep{masson_2006}.

The \citetalias{chandran_intermittency_2015} model predicts $\tilde{\theta}_{\perp}^{z} \propto \lambda^{0.10}$, and $\tilde{\theta}_{\perp}^{ub} \propto \lambda^{0.21}$. For simplicity, in subsequent discussions, when referring to either definition (Equations~\ref{fin_ch_eq:7} and~\ref{fin_ch_eq:8}), $\Theta^{ub}$ will represent magnetic-velocity alignment, and $\Theta^{z}$ will represent Els\"asser alignment angles.

\subsection{Quantifying the Influence of Intermittency and Compressibility on SDDA}\label{subsec:condit_binning_interm_compress}

To assess the influence of intermittency and compressibility on alignment angle measurements, we utilize scale- and time-dependent proxies for these properties. Employing a scale-dependent conditional averaging technique enables us to isolate specific types of fluctuations, thus focusing on particular properties of interest.

To identify coherent structures, we employ the Partial Variance of Increments (PVI) method, which is effective for detecting sharp gradients within a turbulent field \citep{greco_intermittent_2008, servidio_local_2012}. The scale-dependent PVI time series can be estimated as follows:

\begin{equation}\label{fin_ch_eq:9}
    \mathcal{I}_{\boldsymbol{q}} \ = \  \frac{| \delta\boldsymbol{q}( t, \ell)|}{\sqrt{\langle | \delta\boldsymbol{q}( t, \ell)|^{2}\rangle}_{\tau}},
\end{equation}

where $\delta \boldsymbol{q}$ is defined by Equation~\ref{fin_ch_eq:5}, and $\langle...\rangle_{\tau}$ denotes the average over a $\tau = 60~\text{min}$ window \citep{greco_partial_2018}.

To quantify the compressibility of fluctuations, we utilize 

\begin{equation}\label{fin_ch_eq:10}
    n_{\boldsymbol{q}} \ = \left[\frac{\delta |\boldsymbol{q}( t , ~\ell)|}{|\delta \boldsymbol{q}( t ,~ \ell)|}\right]^{2}.
\end{equation}


\par To conduct our analysis, we obtain $\delta \boldsymbol{b}_{\perp}$ and $\delta \boldsymbol{v}_{\perp}$ using the method detailed in Section~\ref{subsec:5pt_increments}. Additionally, we compute $\mathcal{I}_{\boldsymbol{q}}$ and $n_{\boldsymbol{q}}$ using equations~\ref{fin_ch_eq:9} and~\ref{fin_ch_eq:10}, respectively, for a range of $\ell$ values, specifically $\ell = 1.2^{j} d_{i}$, where $j = 15, ..., 80$. 

Here, $d_{i}$ represents the ion inertial length, defined as $d_{i} = { V_{A}}/{\Omega_{i}}$, where $\Omega_{i} = {e|B|}/{m_{p}}$, is the proton gyrofrequency, $e$ is the elementary charge, $|B|$ is the magnetic field magnitude.

\par Subsequently, we employ a scale-dependent conditional binning technique. This process entails partitioning $\mathcal{I}_{\boldsymbol{q}}$, $n_{\boldsymbol{q}}$, and their corresponding alignment angle values into $N= 50$ consecutive bins for each scale $\ell$. More precisely, the $i$-th bin at scale $\ell$ is defined as the interval $[2(i-1), ~2i)$, where $i = 1, ..., 50$. The results of this analysis are presented in Section~\ref{subsec:interm_compress_SDDA}.

\subsection{Quantifying Imbalance Effects on SDDA}\label{subsec:condit_binning_imbalance}

\par The total energy $E_{t} = E^{+} + E^{-}$ and cross-helicity $H_{c} = E^{+} - E^{-}$, expressed in terms of the energy associated with fluctuations in $\boldsymbol{z}^{\pm}$, $E^{\pm} = \langle |\delta \boldsymbol{z}^{\pm}|^{2}\rangle/4$, are ideal (i.e., with zero viscosity and resistivity) invariants of the RMHD equations. Els\"asser imbalance can be quantified by the normalized cross-helicity, $\sigma_c = H_{c}/E_{t}$, which measures the relative fluxes of counterpropagating wavepackets in the system. In the context of solar wind turbulence, Els\"asser imbalance is assessed by examining the relative magnitudes of inwardly and outwardly propagating Alfv\'en waves \citep{Velli_91_waves, Velli_93}.

\begin{equation}\label{fin_ch_eq:12}
\sigma_{c}= \frac{ | \delta \boldsymbol{z}^{o}_{\perp}(t, ~\ell)   |^{2} - |\delta \boldsymbol{z}^{i}_{\perp}(t, ~\ell)   |^{2} }{  | \delta \boldsymbol{z}^{o}_{\perp}(t, ~\ell)   |^{2} + |\delta \boldsymbol{z}^{i}_{\perp}(t, ~\ell)   |^{2} }. 
\end{equation}

\par In addition, we consider the normalized residual energy, $\sigma_r$, to investigate the effects of Alfv\'enic imbalance. This metric evaluates the relative energy in kinetic and magnetic fluctuations:

\begin{equation}\label{fin_ch_eq:13}
\sigma_{r} = \frac{ | \delta \boldsymbol{v}_{\perp}(t, ~\ell)   |^{2} - | \delta \boldsymbol{b}_{\perp}(t, ~\ell)   |^{2}  }{ | \delta \boldsymbol{v}_{\perp}(t, ~\ell)   |^{2}  + | \delta \boldsymbol{b}_{\perp}(t, ~\ell)   |^{2} }.
\end{equation}

\begin{figure*}
     \centering

           \includegraphics[width=1\textwidth]{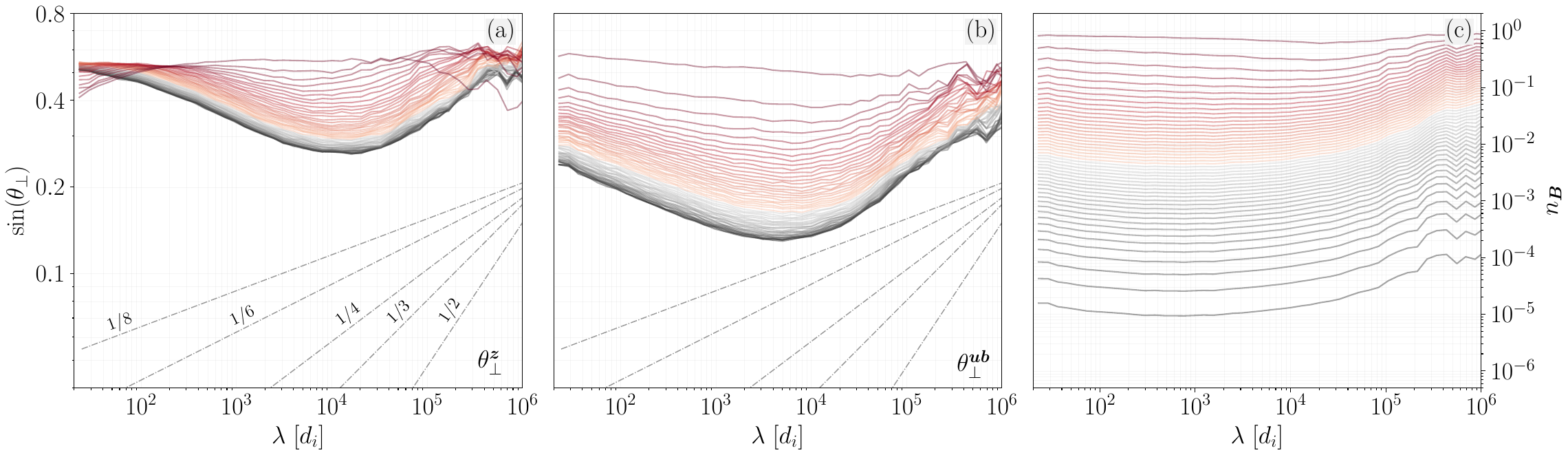}
         \caption{(a) $\sin(\theta_{\perp}^{\boldsymbol{z}})$, (b) $\sin(\theta_{\perp}^{\boldsymbol{ub}})$. Each subplot categorizes alignment angles into 50 bins based on magnetic compressibility, $n_{\boldsymbol{B}}$, with the bins ranging from the lowest magnetic compressibility (black) to the highest magnetic compressibility (red).}
         \label{fig:compress_sdda}
\end{figure*}

\section{Results} \label{sec:Results}

\par To understand the influence of imbalance, we opted for an approach different from the one presented in Section~\ref{subsec:condit_binning_interm_compress}. This decision was based on the results presented Appendix~\ref{appendix:2}. More specifically, among the variables $\sigma_c$, $\sigma_r$, $\theta^{ub}_{\perp}$, and $\theta^{z}_{\perp}$, only two are independent. Thus, using the previously described binning method would merely replicate the relationships shown in Figure~\ref{fig:sig_theta_relationship}. However, a more pressing question remains: how does global-scale imbalance affect SDDA measurements?

\par To address global imbalance, we adhered to the following methodology:
We computed the average values of $\sigma_c(\ell^{\ast})$ and $\sigma_r(\ell^{\ast})$ for each interval identified by visual inspection, where $\ell^{\ast} = 10^{4} ~d_i$. 

\par Following this, we segregate our intervals into 10 linearly spaced bins, ranging from [0, 1] for Els\"asser imbalance and [-1, 0] for Alfv\'enic imbalance, based on the values of $|\sigma_c(\ell_{\ast})|$ and $\sigma_r(\ell^{\ast})$, respectively. Within each bin, we calculate a scale-dependent weighted mean of the alignment angle, using the number of samples in each interval at that scale as the weighting factor. The results of this analysis are discussed in Section~\ref{subsec:imbalance_SDDA}.

In this section, our aim is to assess the impact of intermittency, compressibility, and global imbalance on SDDA measurements. However, before presenting our findings, it is essential to address a key aspect of our analysis.

We first investigated the scaling of SDDA for each stream type (slow Alfv\'enic, fast Alfv\'enic, slow non-Alfv\'enic) separately. Then, we aggregated all homogeneous intervals and applied the conditional averaging method outlined in Section~\ref{subsec:condit_binning_interm_compress}. Interestingly, the trends observed when analyzing stream types with similar characteristics separately persisted when considering a mix of various interval types. This observation could be attributed to the inherently diverse nature of fluctuations across all wind streams. For instance, intervals primarily characterized as incompressible may still contain a fraction of compressible fluctuations. Similarly, wind streams classified as generally balanced may exhibit localized patches of imbalance \citep{Matthaeus_2008_patchy_imbalance, Perez_2009, 2013_Chen}. The use of conditional averaging enables us to concentrate on specific types of fluctuations, facilitating an investigation into particular properties of interest, such as segregating fluctuations based on scale-dependent proxies for compressibility and intermittency.

Finally, it is important to point out that the conditional averaging analysis utilized here only considers normalized quantities like $\theta_{\perp}$, $I_{\boldsymbol{\xi}}$, and $n_{\boldsymbol{\xi}}$. Therefore, at this stage, we did not consider $\tilde{\theta}_{\perp}$, Equation \ref{fin_ch_eq:8}. This is because it entails averaging non-normalized quantities, and extreme values among different intervals, owing to significant variations in the root mean square of the fluctuations, could potentially dominate the mean. 
We will explore the scaling of the alignment angle as defined in Equation \ref{fin_ch_eq:8} in Section~\ref{subsec:imbalance_SDDA}, where weighted averages of the alignment angles  among homogeneous intervals are considered, ensuring that the previously mentioned concern does not hinder our analysis.

\begin{figure*}
     \centering
        \includegraphics[width=1\textwidth]{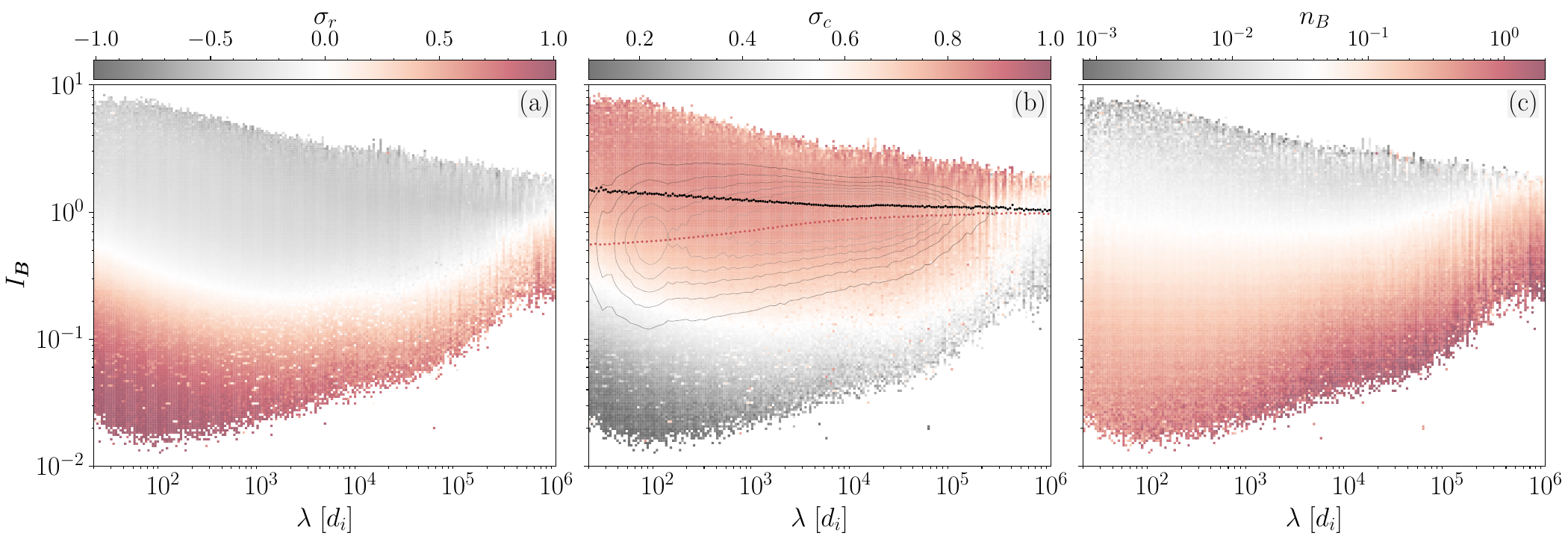}
        \caption{(a) $\sigma_r$. (b) $\sigma_c$. (c) $n_B$ as a function of $\lambda$ and $\mathcal{I}_{\boldsymbol{B}}$. In panel (b), contours indicate the count levels of the distribution. Additionally, panel (b) displays $\mathcal{F}(\mathcal{I}_{\boldsymbol{B}}^{4})^{1/4}$, where $\mathcal{F}$ represents either the mean (black line) or median operator (red line).}
         \label{fig:various_quants}
\end{figure*}

\subsection{Influence of  Intermittency  \&  Compressibility  on measurements of SDDA}\label{subsec:interm_compress_SDDA}

In this section, we delve into the effects of intermittency and compressibility on SDDA using the conditional averaging method described in Section~\ref{subsec:condit_binning_interm_compress}.

Figure~\ref{fig:sdda_pvi} presents the results for intervals of mixed streams. The first two columns show $\theta_{\perp}^{\boldsymbol{z}}$ and $\theta_{\perp}^{\boldsymbol{ub}}$, conditioned on $\mathcal{I}_{\xi}$ with $\xi = \boldsymbol{z}^{o}$, plotted against scales normalized in $d_i$ units. Although analyses for $\xi = \boldsymbol{z}^{i}, \boldsymbol{B}, \boldsymbol{V}$ were also conducted, they are not shown here; key findings are discussed below.

Figure~\ref{fig:sdda_pvi} clearly illustrates that the scaling of SDDA varies significantly depending on the percentile bin considered. At any given scale, an inverse relationship exists between alignment angles and the intensity of field gradients. More specifically, curves derived from averaging alignment angles for fluctuations with weak field gradients (black lines) tend to exhibit flat profiles, indicating negligible or weak SDDA at smaller scales. In contrast, curves for fluctuations with the strongest $\mathcal{I}_{\boldsymbol{z}^{o}}$ indices show steeper scaling laws, suggesting increasing alignment down to $\lambda \approx 8 \cdot 10^{2} d_i$. 

While the anticorrelation holds in both cases, it is apparent that alignment in $\theta_{\perp}^{ub}$ is consistently tighter compared to $\theta_{\perp}^{z}$. This is likely because the intervals considered here are slightly skewed towards the globally Alfv\'enic side, as shown in Figure~\ref{fig:overview}, primarily because such intervals were prioritized during the selection process. This was the only notable difference observed between the analyses of ``pure'' and ``mixed'' intervals. This effect will become more pronounced in Section~\ref{subsec:imbalance_SDDA}, where we demonstrate that $\theta_{\perp}^{ub}$ and $\theta_{\perp}^{z}$ exhibit very similar behavior in balanced intervals, but they tend to diverge, with the former showing tighter alignment as the imbalance increases.  Nonetheless, this does not affect the conclusions of this analysis. The primary objective here is to illustrate the relationship between alignment angles and field gradients, which, as explained earlier, shows consistent behavior in all cases.


For a more straightforward comparison with the numerical simulations by \cite{2015Mallet}, we investigated the relationship between the alignment angle and the normalized Els\"asser fluctuation magnitudes, $|\delta \boldsymbol{z^{o}}|/V_{A}$. The results, not shown here, reveal a trend qualitatively similar to those in Figure~\ref{fig:sdda_pvi}, as expected since the PVI diagnostics provide an estimate of fluctuation gradient intensity normalized by the standard deviation of local gradients.

We proceeded to segregate fluctuations into percentile bins based on the compressibility diagnostic, $n_{\xi}$, as detailed in Section~\ref{sec:data_analysis}. Our findings, illustrated in Figure~\ref{fig:compress_sdda}, reveal a strong correlation between alignment and magnetic compressibility, $n_{\boldsymbol{B}}$. Specifically, at the outer scale, strongly compressible fluctuations exhibit no signs of increasing dynamic alignment; the scaling of SDDA becomes progressively steeper as $n_{B}$ decreases. Conditioning on $n_{\boldsymbol{V}}$ reveals only a weak and statistically insignificant correlation with SDDA, not shown here. It is worth noting, however, that the compressibility in the velocity field is considerably stronger than that in the magnetic field, with $n_{\boldsymbol{V}}(\ell) \gg n_{\boldsymbol{B}}(\ell)$ for the majority of intervals considered.

In general, curves derived from magnetically incompressible fluctuations exhibit behaviors at outer scales qualitatively similar to those associated with the most intense coherent structures, suggesting a strong correlation between compressibility and intermittency \citep[see also][]{Vasko_CS_21, Lotekar_Vasko_21}. However, unlike Figure~\ref{fig:sdda_pvi}, increasing misalignment is observed in most cases at $\lambda \lesssim 10^{4} d_i$, which roughly coincides with the large-scale break separating the energy-injection range from the inertial range of the magnetic field power spectrum, usually associated with $f^{-1}$ and $f^{\alpha}$, $\alpha \in [-5/3, -3/2]$ scalings, respectively \citep[][and references therein]{bruno_solar_2013}.

To quantify our findings, we applied a power-law fit to the curves representing the top 5\% of fluctuations measured by $|\delta \boldsymbol{z}^{\pm}|$ and $\mathcal{I}_{\xi}$, with $\xi = \boldsymbol{z}^{o}, \boldsymbol{z}^{i}, \boldsymbol{B}, \boldsymbol{V}$, and to the bottom 5\% in terms of $n_{\boldsymbol{\xi}}$ for $\xi = \boldsymbol{B}, \boldsymbol{V}$. The derived scalings, evaluated over the range $4 \times 10^{4} - 4 \times 10^{5} d_i$, are summarized in Table~\ref{tab:zp_mag}.

\subsection{Nature of fluctuations}\label{subsec:results_small_scales}

In this section, we aim to better understand the nature of fluctuations at energy-injection and inertial scales. Figure~\ref{fig:various_quants}
 illustrates (a) $\sigma_r$, (b) $\sigma_c$, and (c) $n_B$ as functions of $I_{\boldsymbol{B}}$ and $\lambda$. Additionally, panel (b) includes a contour plot of the $I_{\boldsymbol{B}}$.

The contours of $I_{\boldsymbol{B}}$ show a slight downward trend at inertial scales, indicating a decrease in the median PVI index towards smaller scales. This may seem counterintuitive given in-situ observations of non-Gaussian increment distributions at smaller scales \citep[see e.g.,][]{Sorriso-Valvo99}. However, examining $\mathcal{F}(\mathcal{I}_{\boldsymbol{B}}^{n})^{1/n}$ for $n = 2, \ldots, 6$, where $\mathcal{F}$ represents either the mean (black line) or median operator (red line), reveals that while the median curve trends downward, the mean curve trends upward at scales $\lambda \leq 10^{4} d_i$. The steepness of the mean curve increases with $n$, reflecting intermittency in the magnetic field time series. For $n = 1$, the average is influenced by median-amplitude eddies, while for larger $n$ values, rare large-amplitude eddies dominate. The scale-dependent behavior of both the mean and median suggests a trend towards distributions exhibiting progressively heavier tails at smaller scales, consistent with previous findings \citep{Sorriso-Valvo99, chhiber_subproton}.

\begin{table}
\caption{Power-law scaling of alignment angles within $4\times10^{4} - 4\times10^{5} \, d_i$ using top 5\% fluctuations in $|\delta \boldsymbol{z}^{\pm}|$ and $\mathcal{I}_{\xi}$, where $\xi = \{ \boldsymbol{B}, \boldsymbol{V}, \boldsymbol{z}^{o}, \boldsymbol{z}^{i} \}$, and bottom 5\% in $n_{\boldsymbol{\xi}}$ for $\xi = \{ \boldsymbol{B}, \boldsymbol{V} \}$.}
\label{tab:zp_mag}
\resizebox{\columnwidth}{!}{%
\begin{tabular}{c*{5}{>{$}c<{$}}}
\toprule 
\toprule 
    & \text{$\theta^{z}_{\perp}$}             & \text{$\tilde{\theta}_{\perp}^{z}$}       & \text{$\theta^{ub}_{\perp}$}              & \text{$\Tilde{\theta}_{\perp}^{ub}$}          \\
\midrule 
$\mathcal{I}_{\boldsymbol{z}^{o}}$   & 0.3 \pm 0.02   & 0.32 \pm 0.02 & 0.29\pm 0.02  & 0.33\pm 0.01  \\
$\mathcal{I}_{\boldsymbol{z}^{i}}$   & 0.3 \pm 0.02   & 0.28 \pm 0.03 & 0.30\pm 0.02  & 0.35\pm 0.04   \\
$\mathcal{I}_{\boldsymbol{B}}$       & 0.30 \pm 0.01  & 0.30 \pm 0.02 & 0.27\pm 0.01  & 0.31\pm 0.01    \\
$\mathcal{I}_{\boldsymbol{V}}$       & 0.23 \pm 0.01  & 0.25 \pm 0.02 & 0.29\pm 0.02  & 0.30\pm 0.01     \\
$n_{\boldsymbol{B}}$                 & 0.19 \pm 0.01  & 0.26 \pm 0.02 & 0.25\pm 0.03  & 0.32\pm 0.03     \\
$n_{\boldsymbol{V}}$                 & 0.17 \pm 0.03  & 0.25 \pm 0.03 & 0.33 \pm 0.04 & 0.38 \pm 0.02  \\
\bottomrule 
\end{tabular}}
\end{table}

Conversely, the roughly scale-independent behavior observed in both the mean and median curves at larger scales reflects the well-established phenomenon of Gaussian distributions of fluctuations at energy injection scales \citep{Sorriso-Valvo99, bruno_radial_2003}.

Interestingly, the transition from intermittent to Gaussian regimes appears to coincide with the scale at which we typically observe the switch from increasing alignment to increasing misalignment, as shown in Figure~\ref{fig:compress_sdda}. However, the impact of this regime change on the scaling of SDDA remains unclear.

Furthermore, we observe that at any given scale, fluctuations with the highest PVI index exhibit the lowest $n_B$ and highest $\sigma_c$ values, along with strongly negative residual energy indices. This implies an anticorrelation between $\sigma_r$ and $I_{\boldsymbol{B}}$, consistent with the findings of \cite{2018_Bowen_residual}, who noted that intervals with large scale-dependent kurtosis, $K(\ell)$, predominantly exhibit negative $\sigma_r$. Our analysis confirms the persistence of this relationship across the observed spectrum, substantiating the claim that intermittency and residual energy are most likely interconnected.

\begin{figure}[t]
     \centering
     \includegraphics[width=0.45\textwidth]{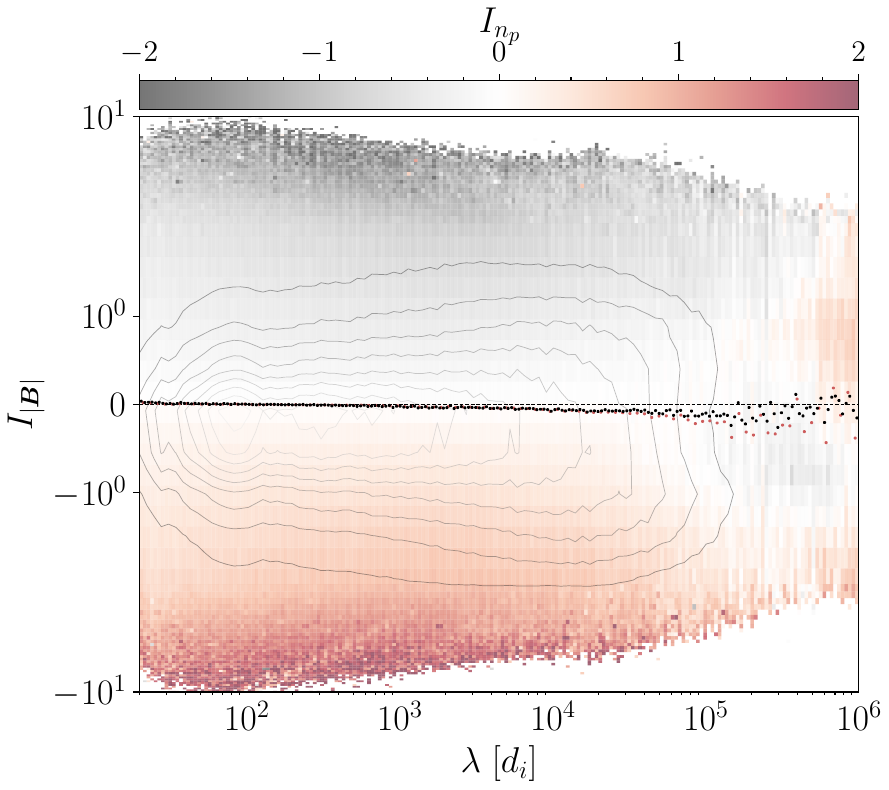}
     \caption{$I_{n_p}$, as a function of $I_{|B|}$ and scale $\lambda$ (normalized to the ion inertial length $d_i$). The plot utilizes a color gradient to represent the mean values of $I_{n_p}$ within each bin. Black and red dots show the mean and median values of $I_{|B|}$ as a function of scale.}
     \label{fig:pvi_mod}
\end{figure}

Finally, we turn our attention to the nature of the compressible fluctuations. To this end, we consider the PVI diagnostic applied to the magnetic field magnitude and proton density time series, $I_{|\boldsymbol{B}|}$ and $I_{n_p}$, respectively. Figure~\ref{fig:pvi_mod}
 illustrates $I_{n_p}$ as a function of $I_{|B|}$ and scale $\lambda$. This method of visualization has not been previously implemented. Earlier works have compared the magnetic and thermal pressure \citep{Burlaga_1970, Thieme_1990}, used wavelet cross-coherence analysis \citep{Kellogg_2005, Yao_2011_B_np_anticor}, or utilized the zero-lag cross-correlation between $|B|$ and $n_p$ \citep{Howes_2012, 2018_Bowen}.

A well-established result is confirmed: fluctuations in $|B|$ and $n_p$ are anticorrelated at inertial scales, similar to observations by \citet{Thieme_1990, Yao_2011_B_np_anticor, howes_slow-mode_2012, 2018_Bowen}, attributed to either the presence of the kinetic slow-mode \citep{howes_slow-mode_2012, Klein_2012} or the MHD slow-mode \citep{Verscharen_2017}. Focusing on the strongly compressible fluctuations, we observe that the anticorrelation persists at the outer scale, $\lambda \gtrsim 10^{4} d_i$, with a tendency for both $I_{n_p}$ and $I_{|B|}$ to predominantly obtain negative values, indicating a preference for magnetic field magnitude depressions accompanied by a decrease in proton density. Finally, at even larger scales, $\lambda \geq 5 \times 10^{5} d_i$, a rough correlation is observed, consistent with the results of \cite{Burlaga_1970}.

It is not clear whether this observation can provide any insights into the effects of compressible fluctuations on SDDA. Nevertheless, we present this finding for completeness without further discussion.

\begin{figure*}
     \centering

           \includegraphics[width=1\textwidth]{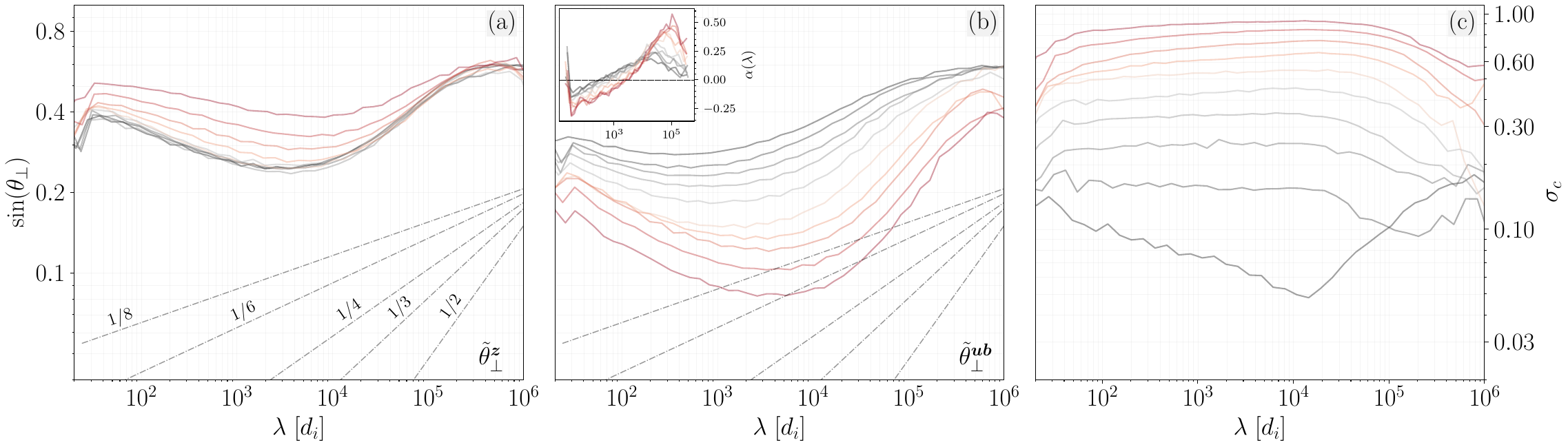}
         \caption{Weighted averages of (a) $\sin(\Tilde{\theta}_{\perp}^{\boldsymbol{z}})$, (b) $\sin(\Tilde{\theta}_{\perp}^{\boldsymbol{ub}})$, and (c) $\sigma_{c}$ for the homogeneous intervals identified through visual inspection. After estimating these quantities as a function of scale for each identified interval, the curves are divided into $N=10$ bins based on $\sigma_{c}(\ell^{\ast})$, where $\ell^{\ast} = 2\cdot 10^{4} d_i$, and a scale-dependent weighted average of the curves is computed.
}
         \label{fig:SDDA_sig_c}
\end{figure*}

\subsection{Influence of Imbalance on SDDA}\label{subsec:imbalance_SDDA}

In this section, we seek to understand the influence of global Alfv\'enic and Els\"asser imbalance on alignment angle scaling using homogeneous intervals identified through visual inspection (see Section~\ref{sec:Data}). To this end we employ the conditional averaging technique described in Section~\ref{subsec:condit_binning_imbalance}, and segregate intervals based on the outer scale values of either the  cross-helicity or normalized residual energy $\sigma_c(\ell^{\ast})$ and $\sigma_r(\ell^{\ast})$, where $\ell^{\ast} = 10^{4}d_i$.

We first consider alignment curves conditioned on $\sigma_c(\ell^{\ast})$. The results are shown in Figure~\ref{fig:SDDA_sig_c}. For $\sigma_c(\ell^{\ast}) < 0.7$, there is no clear trend in the scaling of $\tilde{\theta}_{\perp}^{z}$, as curves from all bins overlap. However, for $\sigma_c(\ell^{\ast}) > 0.7$, the scaling of $\tilde{\theta}_{\perp}^{z}$ becomes progressively shallower, especially at larger scales. In contrast, large-scale Els\"asser imbalance organizes the $\tilde{\theta}_{\perp}^{ub}$ curves more clearly. As $\sigma_c(\ell^{\ast})$ increases, (a) $\tilde{\theta}_{\perp}^{ub}$ shows steeper scaling, and (b) the rollover of the curves, i.e., the range at which increasing misalignment is observed, shifts towards larger scales. The local scaling index, $\alpha(\lambda)$, illustrated in the inset of Figure~\ref{fig:SDDA_sig_c}b, shows that $\alpha(\lambda) \approx 0.4$ for $\sigma_c(\ell^{\ast}) > 0.90$, while a shallower scaling of $\alpha(\lambda) \approx 0.2$ is observed for $\sigma_c(\ell^{\ast}) < 0.10$. The range of scales over which the slope remains positive increases as $\sigma_c(\ell^{\ast}) \rightarrow 0$.

It is worth noting that while alignment in $\tilde{\theta}_{\perp}^{ub}$ is considerably tighter compared to $\tilde{\theta}_{\perp}^{z}$ for strongly imbalanced intervals, the two definitions yield qualitatively similar results at $\sigma_c(\ell^{\ast}) \rightarrow 0$.

When intervals are segregated based on $\sigma_r(\ell^{\ast})$, a similar trend is observed for $\tilde{\theta}_{\perp}^{ub}$. Additionally, as Alfv\'enic imbalance decreases ($\sigma_r(\ell^{\ast}) \rightarrow 0$), SDDA signatures in $\tilde{\theta}_{\perp}^{z}$ diminish, resulting in flatter curves between $10^{2}-10^{6} d_i$. Conversely, with increasing magnetic energy dominance over kinetic energy, SDDA signatures become more pronounced, exhibiting steep power-law behavior, particularly at larger scales. In practice, this implies that equipartitioned intervals ($E_{u} \approx E_{b}$) show no signs of increasing alignment in the Els\"asser field fluctuations, but at least at large scales still show increasing alignment in the polarizations of magnetic/velocity field fluctuations.

To summarize:
(a) Large-scale Alfv\'enic imbalance is a good predictor for the expected behavior of both $\tilde{\theta}_{\perp}^{ub}$ and $\tilde{\theta}_{\perp}^{z}$. As $\sigma_r(\ell^{\ast})$ approaches zero, the scalings of $\tilde{\theta}_{\perp}^{ub}$ become steeper, while the scalings of $\tilde{\theta}_{\perp}^{z}$ become shallower, extending to a narrower and wider range of scales, respectively.
(b) Els\"asser imbalance mostly correlates with the $\tilde{\theta}_{\perp}^{ub}$ scaling. SDDA signatures persist to smaller scales for globally balanced streams compared to their imbalanced counterparts, but with the scaling becoming progressively shallower. Is it worth noting that the persistence of SDDA signatures to smaller scales is consistent with noise being responsible for misalignment at small scales. That is because the globally imbalanced streams reach small alignment angles at larger scales and so become affected by noise at larger scales.

\begin{figure*}
     \centering
           \includegraphics[width=1\textwidth]{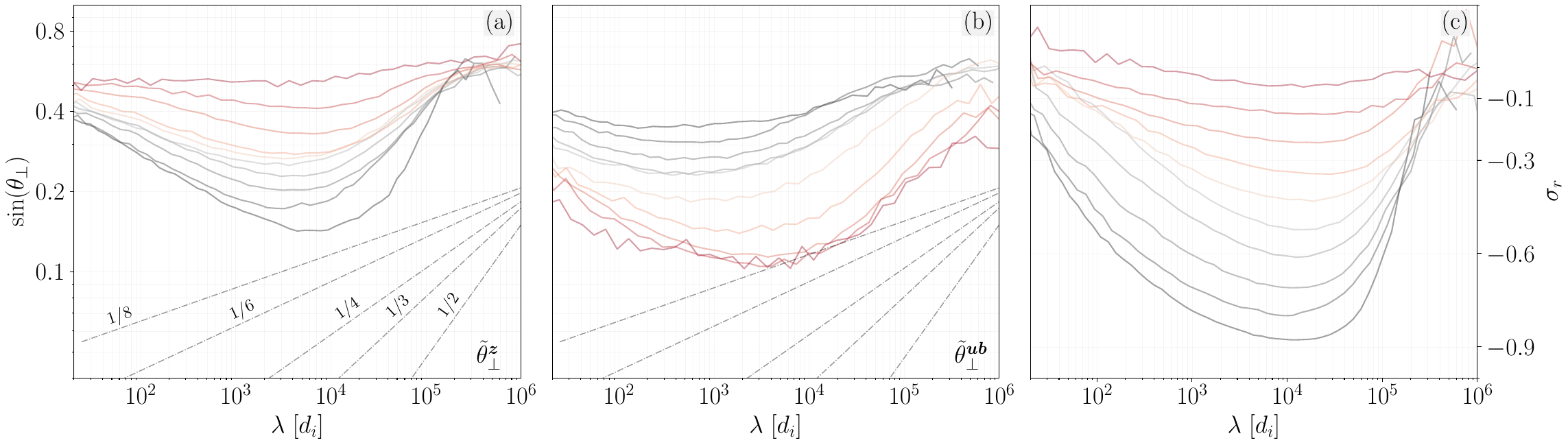}
         \caption{Weighted averages of (a) $\sin(\Tilde{\theta}_{\perp}^{\boldsymbol{z}})$, (b) $\sin(\Tilde{\theta}_{\perp}^{\boldsymbol{ub}})$, and (c) $\sigma_{r}$ for the homogeneous intervals identified through visual inspection. After estimating these quantities as a function of scale for each identified interval, the curves are divided into $N=10$ bins based on $\sigma_{r}(\ell^{\ast})$, where $\ell^{\ast} = 2\cdot 10^{4} d_i$, and a scale-dependent weighted average of the curves is computed.}
         \label{fig:SDDA_sig_r}
\end{figure*}

\subsection{High-Frequency Instrumental Noise}\label{subsec:instr-noise-modeling:theory}

In this section, we investigate the extent to which high-frequency, small-amplitude noise in the velocity field measurements can affect our ability to reliably estimate alignment angles in the Els\"asser field fluctuations, $\theta_{\perp}^{z}$.

Let $\delta \boldsymbol{b}$ denote the true value of the magnetic field fluctuation vector in velocity units. We assume that the error in its measurement is negligible. Although a more realistic assessment would incorporate an error term, especially since $\delta \boldsymbol{b}$ measurements may be contaminated by errors in proton density measurements, we assume for highly Alfv\'enic intervals that $\delta \rho / \rho$ is sufficiently small. This assumption is supported by observations \citep{shi_alfvenic_2021}, allowing rolling averages of $\rho$ to be used for the  Alfv\'enic normalization.

We define the real value of the velocity field as $\delta \boldsymbol{u}$ and introduce an error term $\boldsymbol{\Delta}$, where $\boldsymbol{\Delta}$ is a normally distributed random vector with mean $\boldsymbol{0}$ and standard deviation $\sigma$, represented by $\boldsymbol{\Delta} \sim \mathcal{N}(\boldsymbol{0}, \sigma)$. We assume that the magnitude of $\boldsymbol{\Delta}$ is significantly smaller than the magnitudes of $\delta \boldsymbol{b}$ and $\delta \boldsymbol{u}$ (i.e., $|\boldsymbol{\Delta}| \ll |\delta \boldsymbol{b}|, |\delta \boldsymbol{u}|$). For simplicity, we assume that the real vectors are perfectly aligned, $\delta \boldsymbol{u} = \alpha \delta \boldsymbol{b}$, and that $\delta \boldsymbol{z}_{i} = \epsilon \delta \boldsymbol{z}_{o}$.

We can then write:

\begin{equation}
\delta \boldsymbol{z}^{\prime}_{o} = \delta \boldsymbol{z}_{o} + \boldsymbol{\Delta}
\end{equation}

\begin{equation}
\delta \boldsymbol{z}_{i}^{\prime} = \epsilon \delta \boldsymbol{z}_{o} + \boldsymbol{\Delta}.
\end{equation}

The quantity controlling the error in the measurements of $\delta \boldsymbol{z}_{i}$ can be represented by the dimensionless parameter $Q$:

\begin{equation}
Q = \frac{\epsilon \delta z_{o}}{\Delta}
\end{equation}

We can estimate the alignment angles as follows:

\begin{align*}
\cos(\theta^\prime) &= \frac{\delta 
 \boldsymbol{z}_{o}^{\prime} \cdot \delta 
 \boldsymbol{z}_{i}^{\prime}}{|\delta 
 z_{o}^{\prime}| |\delta 
 z_{i}^{\prime}|} \\
&= \frac{\epsilon \delta  z_{o}^{2} + (1+\epsilon) \delta  \boldsymbol{z}_{o} \cdot \boldsymbol{\Delta} + \Delta^{2}}
{\epsilon \delta  z_{o}^{2} \sqrt{1 + \frac{2 \delta  \boldsymbol{z}_{o} \cdot \boldsymbol{\Delta}}{\delta  z_{o}^{2}} + \frac{\Delta^{2}}{z_{o}^{2}}} \sqrt{1 + \frac{2 \delta  \boldsymbol{z}_{o} \cdot  \boldsymbol{\Delta}}{\epsilon \delta  z_{o}^{2}} + \frac{\Delta^{2}}{{\epsilon}^{2} \delta 
 z_{o}^{2}}}}
\end{align*}

Considering the case where $Q \gg 1$, we can estimate:

\begin{equation}
\cos(\theta^{\prime}) \approx 1 - O(Q^{-1}) 
\end{equation}

On the other hand, when $Q \ll 1$:


\begin{equation}
\cos(\theta^{\prime}) \approx\delta \boldsymbol{z}^o \cdot \boldsymbol{\Delta}/(\delta z^o \Delta).
\end{equation}
Therefore, the accuracy of the angle measurement depends on the errors in the velocity compared to the true size of the $\delta z_{i}$ signal. Specifically, when $Q \gg 1$, the alignment angle can be estimated reliably. Conversely, for $Q \ll 1$, a significant deviation is expected between the true and estimated values of the alignment angles.

\subsection{Modeling High Frequency Instrumental Noise: Perfectly aligned fields.}\label{subsec:instr-noise:modeling:1}

To further investigate the impact of high-frequency noise on alignment angle measurements, we simulate the previous analysis by generating synthetic magnetic and velocity field data, subsequently polluted by artificial high-frequency white noise of specified amplitude. The synthetic time series, power spectra, and alignment angle estimates are obtained following the method described in Appendix~\ref{appendix:synthetic-timeseries-aligned}.

\begin{figure*}
     \centering

           \includegraphics[width=1\textwidth]{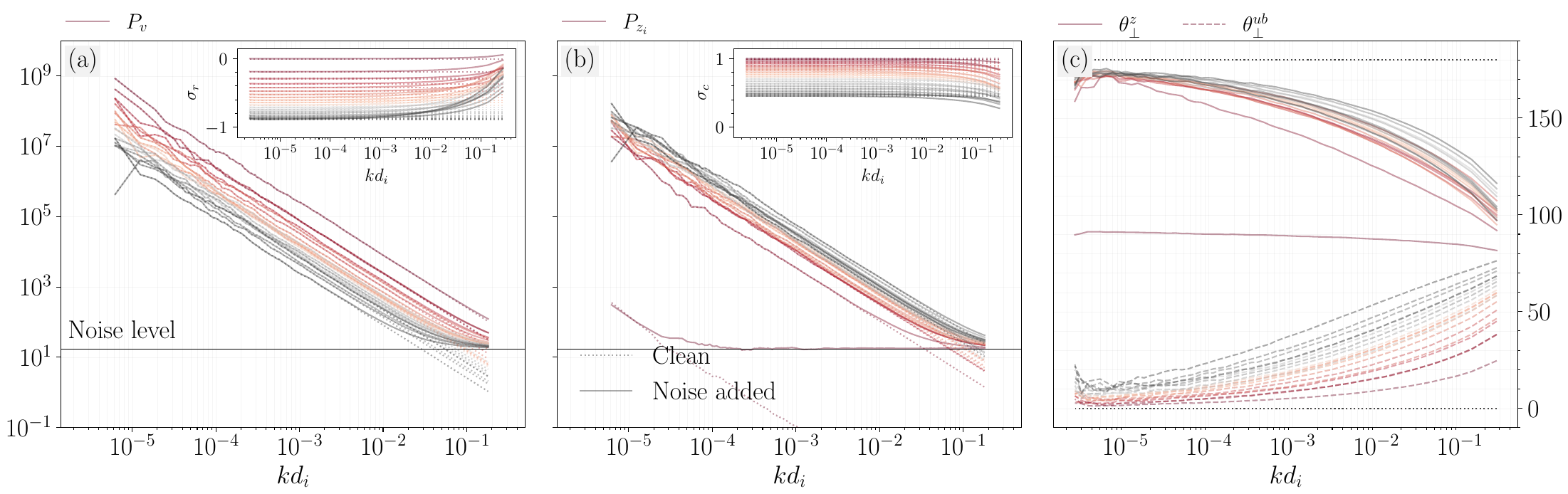}
         \caption{Synthetic power-spectra of (a) the velocity field ($P_{\boldsymbol{v}}$) and (b) the minor Els\"asser field ($P_{\boldsymbol{z}_{i}}$) for different levels of imbalance. Dotted lines indicate the original time series, while solid lines represent the time series after adding white noise. Panel (c) shows the alignment angle of the Els\"asser/Alfv\'en fields for the original time series (dotted lines), which remain constant at $\theta^{z} = 180^{\circ}$, $\theta^{ub} = 0^{\circ}$, respectively, independent of scale, and for the noise-affected time series (solid lines). The insets in panels (a), (b) illustrate $\sigma_c$, $\sigma_r$ for the different runs, respectively.}
         \label{fig:simulate_noise_level}
\end{figure*}

The results of this analysis are illustrated in Figure~\ref{fig:simulate_noise_level}. Panels (a)-(c) display the power-spectral density of the velocity field $P_{\boldsymbol{v}}$, the ingoing Els\"asser fields $P_{z_{i}}$, and the alignment angles $\theta_{\perp}^{z}$ and $\theta_{\perp}^{ub}$, respectively. The insets in panels (a) and (b) show $\sigma_r$ and $\sigma_c$. Different colors represent varying levels of Els\"asser imbalance, with red lines indicating imbalanced intervals and black curves showing balanced intervals. Dotted lines represent the expected behavior, while solid lines show the impact of introducing white noise into the velocity-field time series.

Introducing noise causes $P_{\boldsymbol{v}}$ to flatten as $k d_i$ increases beyond a critical value $k_n d_i \approx 10^{-2}$ for the lowest cross-helicity intervals, as shown in Figure~\ref{fig:simulate_noise_level}(a). As $\sigma_c \rightarrow 1$, $\boldsymbol{V}_{RMS}$ increases and becomes comparable to $\boldsymbol{B}_{RMS}$. This results in a decrease in $\boldsymbol{z}_{i,\text{RMS}}$, leading to an artificial flattening of the power spectrum of $\boldsymbol{z}_{i}$ at progressively smaller scales. Consequently, in the highest $\sigma_c$ bin, $P_{z_{i}}$ flattens at a $k d_i \ll k_n d_i$.

When $P_{\boldsymbol{v}}$ reaches the noise floor at $k_n d_i$, the power spectrum of $\boldsymbol{z}_{i}$ becomes unmeasurable for $P_{z_{i}}(k d_i > k^{\ast} d_i)$, where $P_{z_{i}}(k^{\ast} d_i) = P_{u}(k_n d_i)$. For the balanced intervals (black curves), although $P_{\boldsymbol{v}}$ is flattened at smaller scales, the impact on $P_{z_{i}}$ is subtle and mainly evident at large scales.

Although our assumption that the noise floor is independent of imbalance might be an oversimplification, our analysis demonstrates that small amplitude high-frequency noise can significantly affect $P_{z_{i}}$, thereby reducing the accuracy of $P_{z_{i}}$ estimates as the Els\"asser imbalance increases.

Noise has an even greater impact on alignment angle estimates. It distorts the phases of the velocity field measurements, causing them to decorrelate from the magnetic field phases without significantly altering their amplitude. The extent and scale at which these effects become significant depend on the level of imbalance: $\theta_{\perp}^{z}$ is more affected during imbalanced intervals, while $\theta_{\perp}^{ub}$ is more affected during balanced intervals. This can be attributed to the fact that the RMS of the velocity field fluctuations is considerably lower for balanced intervals (see, e.g., Figure \ref{fig:all_fields_RMS}), and thus the noise floor is reached at smaller scales.

It is important to note that by imposing perfect alignment between magnetic and velocity field fluctuations, simulating different levels of imbalance requires the RMS of the velocity field fluctuations to decrease faster than that of the magnetic field fluctuations for $\sigma_c$ to decrease. The opposite scenario could also be considered, but that would result in positive $\sigma_r$, which is not usually observed in the solar wind \citep[][and references therein]{bruno_solar_2013}. With this assumption, we have effectively restricted our analysis to the periphery of the circle plot shown in Figure \ref{fig:sig_theta_relationship}. However, this is not always the case; in principle, we could maintain the RMS of both fields constant and adjust the alignment angles or use a combination of these factors. Given that the ratio of the RMS in the synthetic time series is informed by our in-situ observations, while the correlation in the fluctuations of the fields is not, our approach does not allow $\sigma_c$ at large scales to decrease below $\sigma_c \lesssim 0.45$ due to our assumption of perfect alignment.

\subsection{Modeling High Frequency Instrumental Noise: Correlation coefficient informed by in-situ observations.}\label{subsec:instr-noise:modeling:2}

To address the simplifications made in Section~\ref{subsec:instr-noise:modeling:1}, we generate time series for the magnetic and velocity fields that satisfy three simultaneous conditions based on our in-situ observations: (1) the fluctuations are correlated with a specified correlation coefficient, (2) their inertial-range spectral scaling matches the observed scaling, and (3) the RMS values of these time series align with those observed in-situ. A detailed description of the steps followed to generate the synthetic data is presentd in Appendix~\ref{appendix:synthetic-timeseries-non- aligned}.

\begin{figure*}
     \centering
           \includegraphics[width=1\textwidth]{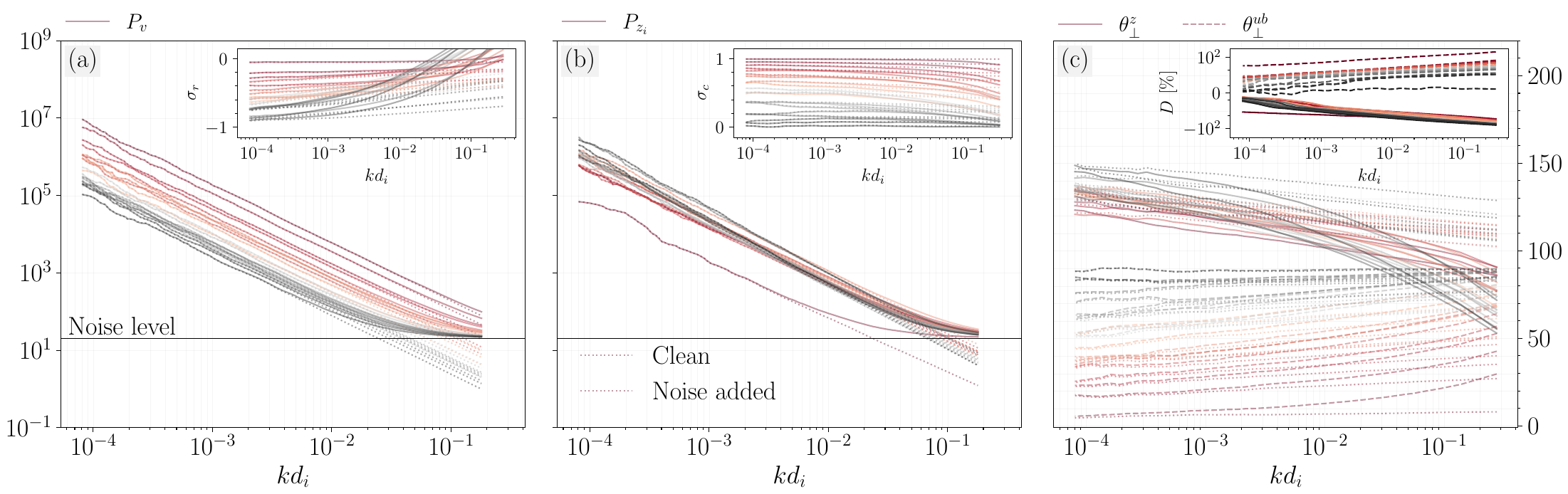}
         \caption{The setup is similar to that in Figure~\ref{fig:simulate_noise_level}, but with the time series generated using the empirical relationships obtained for $\rho$, $\boldsymbol{B}_{\text{RMS}}$, $\boldsymbol{V}_{\text{RMS}}$, $a_{b}$, and $a_{u}$. Additionally, the inset in panel (c) features the normalized difference $D = 100 (\theta - \theta_c)/ \theta_c$, where $\theta$ and $\theta_c$ are the noise-affected and clean estimates of the alignment angles, respectively.}
         \label{fig:simulate_noise_level_2}
\end{figure*}

The results of this analysis are presented in Figure~\ref{fig:simulate_noise_level_2}, with a setup identical to Figure~\ref{fig:simulate_noise_level}. In addition the inset of panel (c) illustrates the normalized diffrence of the alignemnt angle estimates of the clean and noise-affected timeseries. While the exact spectral scaling of the different fields does not make a big difference in the observed behavior, not shown here, allowing the correlation coefficient, to vary in a manner consitent with in-situ data, enables for a more realistic comparison between the synthetic and in-situ data. 

To further clarify this aspect, we estimate the band-pass RMS value of the different fields within a specified $k^{\ast} = kd_i$, range from the power spectral density through:

\begin{equation}\label{fin_ch_eq:band_pass_RMS}
\mathcal{J}_{\boldsymbol{\phi}} = \sqrt{\int_{k^{\ast}_{j}}^{k^{\ast}_{j+1}} P_{\boldsymbol{\phi}}(k^{\ast}) \, dk^{\ast}}
\end{equation}
where, $k^{\ast}_{j}$ and $k^{\ast}_{j+1}$ denote the lower and upper bounds of the $j$-th  bin, respectively, with $k^{\ast}_{j} \in [10^{-4}, 4 \times 10^{-1}]$. The bins are linearly spaced in logarithmic space.

\begin{figure*}
     \centering
           \includegraphics[width=1\textwidth]{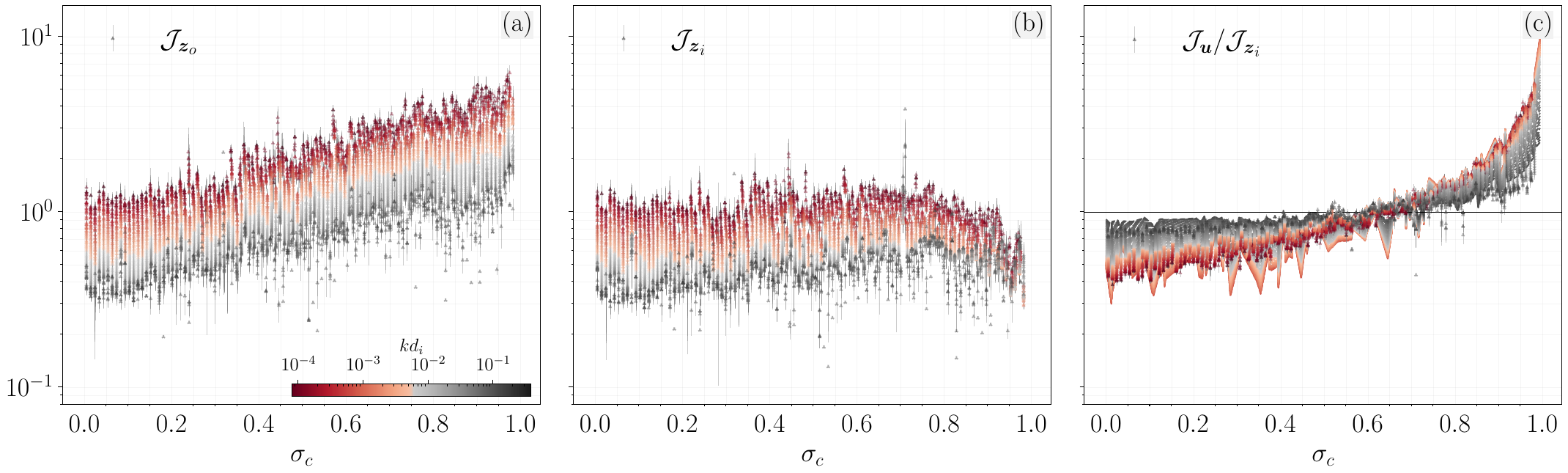}
            \caption{(a) $\mathcal{J}_{\boldsymbol{z}_{o}}$, (b) $\mathcal{J}_{\boldsymbol{z}_{i}}$, and (c) $\mathcal{J}_{\boldsymbol{v}} / \mathcal{J}_{\boldsymbol{z}_{i}}$ as functions of $\sigma_c$. Here, $\mathcal{J}_{\boldsymbol{\phi}}$ represents the band-pass RMS of the field $\boldsymbol{\phi}$, as defined in Equation \ref{fin_ch_eq:band_pass_RMS}. These values are initially estimated for individual intervals in our dataset. Subsequently, the weighted mean values at a given $kd_i$ and $\sigma_c$ bin are calculated, with the weights being the interval duration.  Panel (c) also includes the estimates of the same quantities for the synthetic dataset obtained by integrating the power-spectra presented in Figure \ref{fig:simulate_noise_level_2}. These are shown as solid lines overlaid on the dots from the in-situ data.}
         \label{fig:J_quant_sig_c}
\end{figure*}

The results of this analysis are presented in Figure~\ref{fig:J_quant_sig_c}. Panels (a), (b), and (c) illustrate the variations of $\mathcal{J}_{\boldsymbol{z}_{o}}$, $\mathcal{J}_{\boldsymbol{z}_{i}}$, and $\mathcal{J}_{\boldsymbol{v}} / \mathcal{J}_{\boldsymbol{z}_{i}}$ as functions of $kd_i$ and $\sigma_c$. These values are initially calculated for individual intervals in our dataset. Afterward, the weighted mean values for each $kd_i$ and $\sigma_c$ bin are determined, with the weights being the interval duration.

Panel (c) additionally includes estimates of the same quantities derived from the synthetic dataset, obtained by integrating the power-spectra shown in Figure \ref{fig:simulate_noise_level_2}. These synthetic estimates are represented as solid lines overlaid on the dots from the in-situ data. 

The good agreement between the synthetic and in-situ data, at all scales considered, suggests that the noise level introduced in the synthetic dataset is roughly consistent with the quantization noise observed in the in-situ data.

Overall, this analysis indicates that small amplitude high-frequency noise can cause significant deviations in observed alignment angles from their true values, even at low frequencies (large scales). However, reliable estimates of $\sigma_c$ and $\sigma_r$ can still be obtained at higher frequencies, as the distortion in amplitude is minimal, but the phase effects are significant.


\section{Discussion}\label{sec:Disussion}

In recent years, the inertial-range scaling behavior of the alignment angle has ignited intense discussions and debate. SDDA, if related to the reduction of nonlinearities\footnote{A dedicated discussion on why this may not be the case can be found in \cite{bowen2021nonlinear}}, holds the potential to flatten the inertial range spectrum from $E(k_{\perp}) \propto k_{\perp}^{-5/3}$ to $E(k_{\perp}) \propto k_{\perp}^{-3/2}$. Although the spectral exponents of -5/3 and -3/2 are numerically close, they signify fundamentally different physical mechanisms underlying the turbulent energy cascade. Therefore, a thorough understanding of the inertial range behavior of SDDA is crucial, particularly when applying MHD turbulence models to astrophysical systems with extensive inertial ranges. For example,  predictions about the turbulent heating rate can deviate significantly between models that ingore and incorporate SDDA  \citep{Chandran_2010, Chandran_Perez_2019}.

Nonetheless, in-situ observations have cast doubt on our comprehension of the role of SDDA in Alfv\'enic turbulence. While several aspects of model incorporating SDDA \citep{boldyrev_2006, Chandran_2015, Mallet_2017} are consistent with in-situ observations \citep{Chen_2012ApJ, Verdini_3D_2018, 2023_Chen_compres, sioulas2024higherorder}, signatures of increasing alignment usually fade within the inertial range \citep{Podesta_2009, Verdini_3D_2018, Parashar_2019, Parashar_2020, sioulas2024higherorder}.

It is imperative to recognize, however, that the \citetalias{boldyrev_2006} and \citetalias{chandran_intermittency_2015} models omit the potential effects of compressibility, imbalance, solar wind expansion, and various instabilities on field alignment. These factors have been deliberately neglected either because they were deemed negligible, given that these models focus on homogeneous alfv\'enic turbulence, or to simplify computations, e.g., by assuming negligible cross-helicity.  Nonetheless, these effects often become significant in the solar wind, raising questions about the applicability of these models to such conditions \citep{Verdini_3D_2018, bowen2021nonlinear}.

In this work, we have cast our analysis within a framework that allows us to isolate and study the influence of effects such as compressibility, intermittency, and imbalance, which have not been adequately addressed in previous works.

Before drawing conclusions in Section~\ref{sec:Conclusions}, we review related work and discuss the characteristics of solar wind turbulence that diverge from conventional models of homogeneous MHD turbulence. Additionally, we evaluate how instrumental noise might obscure the precise estimation of alignment angles in solar wind studies.

\subsection{Wavepackets, shearing, alignment \& residual energy}\label{subsec:shearing}

We have shown that the quenching of the alignment angle typically observed at inertial scales shifts toward smaller scales when fluctuations with strong gradients are isolated from residual fluctuations,  Figure~\ref{fig:sdda_pvi}a. It is well known that the breadth of the inertial range narrows with Els\"asser imbalance \citep[][and references therein]{bruno_solar_2013}. Nevertheless, at 1 AU, the inertial range begins, on average, at $\ell \approx 10^4 d_i$. Combining this observation with the results presented in Figure~\ref{fig:sdda_pvi}, it is clear that by carefully thresholding based on the field gradients, SDDA signatures can still be detected across a significant portion of the inertial range.

This latter observation is crucial because it is not the typical or median amplitude eddy that is expected—or required for the purposes of the \citetalias{chandran_intermittency_2015} model—to be strongly aligned at a given scale. Such wave packets have likely undergone several balanced collisions, which could impede increasing alignment. Instead, it is the ``atypical,'' yet ``dynamically relevant,'' eddies residing at the tails of the PDFs of increments that are expected to show such behavior for the model to work. This is indeed the case, at least down to the point where instrumental effects likely dominate the statistics (see analysis in Section~\ref{subsec:instr-noise:modeling:2}). These wave packets are characterized by strong field gradients, remain strongly incompressible, and have highly correlated velocity/magnetic field fluctuations, as evidenced by the low $n_{B}$ and high $\sigma_c$ indices, respectively.  In this sense, our results are consistent with the CSM15 model, although it should be noted that CSM15 did not account for the residual energy that we and others find to be widespread in solar-wind turbulence.\footnote{It is perhaps worth noting that although the average cross helicity was zero in the \citetalias{chandran_intermittency_2015} model, the local cross helicity was high within the volume occupied by the large-amplitude fluctuations in the tail of the distribution.}

Taking all this into account, our observations support the following picture: The nonlinear, intermittent dynamics perpendicular to $\boldsymbol{B}_{0}$ produce highly aligned, strongly turbulent, current sheet structures that retain certain properties associated with the linear-wave response, given their high cross-helicity. Kinetic-magnetic energy equipartition is spontaneously broken due to imbalanced wave packet collisions, even if it is present initially at large scales, where $\delta \boldsymbol{u} \approx \pm \delta \boldsymbol{b}$. This process culminates in the formation of incompressible current sheet structures at smaller scales, residing at the periphery of the circle plot illustrated in Appendix~\ref{appendix:2}, a result also recovered in homogeneous and EBM simulations \citep[see e.g.,][]{Wang_residual_2011, Dong_2014, 2023_Chen_compres, meyrand2023reflectiondriven}. As such, our results provide observational support for the in-situ generation of coherent structures as a by-product of the turbulent cascade. At the same time, however, our results cannot definitively rule out the presence of advected flux tube structures originating from the inner heliosphere \citep{borovsky_flux_2008}.

\subsection{Dependence of SDDA measurements to Els\"asser and Alfv\'enic imbalance}\label{subsec:discuss_imbalancee}

In Section~\ref{subsec:imbalance_SDDA}, we demonstrated that the scaling of $\Theta^{z}$ and $\Theta^{ub}$ is strongly correlated with the global Alfv\'enic and Els\"asser imbalance, but with some nuances.

For instance, the large-scale value of normalized cross helicity, $\sigma_c(\ell^{\ast})$, with $\ell^{\ast} = 2 \times 10^{4} d_i$, shows a strong correlation with $\Theta^{ub}$. As $\sigma_c(\ell^{\ast}) \rightarrow 1$, the scaling of $\Theta^{ub}$ becomes steeper and is observed over a progressively narrower range of scales. However, the scaling of $\Theta^{z}$ does not appear to change significantly with global Els\"asser imbalance, at least for $\sigma_c(\ell^{\ast}) < 0.7$. This finding is in qualitative agreement with the incompressible, homogeneous MHD simulations by \citet{Beresnyak_2009_imbalanced}, who found that the scaling of $\Theta^{z}$ remains unaffected by the global Els\"asser imbalance.

Conversely, large-scale Alfv\'enic imbalance correlates with both $\Theta^{z}$ and $\Theta^{ub}$. As $\sigma_r(\ell^{\ast}) \rightarrow 0$, the scaling of $\Theta^{ub}$ becomes steeper and narrows in scale range; that is, the rollover of the curves shifts to smaller scales as $\sigma_r(\ell^{\ast}) \rightarrow -1$ or $\sigma_c(\ell^{\ast}) \rightarrow 0$. For $\Theta^{z}$, the scaling becomes steeper as $\sigma_r(\ell^{\ast}) \rightarrow -1$, but there is no clear trend in the extent of $\Theta^{z}$ with $\sigma_r(\ell^{\ast})$.

While our analysis clearly indicates that the scaling of SDDA depends strongly on the degree of large-scale imbalance in the system, the observation that the extent over which these signatures persist might also depend on imbalance is less convincing. It is tempting to assume that for globally imbalanced intervals, the fields are already tightly aligned at large scales, resulting in a strong depletion of nonlinearities. This would effectively halt the mechanism that results in SDDA and prevent the average value of the different alignment angles from becoming any tighter towards smaller scales. This could lead to a saturation of the alignment angles at some minimum value, or allow for other mechanisms that result in misalignment of the fields to become significant, as discussed in Section~\ref{subsec:discuss_instabilities}. 

Another possible explanation could be instrumental effects. As discussed in Section~\ref{subsec:instr-noise:modeling:2}, small uncertainties in high-frequency velocity field measurements can result in significant uncertainty in alignment angle measurements, with the effect becoming increasingly important as Els\"asser imbalance increases. For example, it is clear from the inset in Figure~\ref{fig:simulate_noise_level_2}c that as $\sigma_c \rightarrow 1$, the effects of high-frequency noise become increasingly important at larger scales for $\Theta^{ub}$, consistent with the results presented in Figure~\ref{fig:SDDA_sig_r}b. Thus, the dependence of alignment angle contamination due to instrumental effects on the imbalance could provide an alternative explanation for the observed trend.

In any case, the extent to which SDDA persists deeper into the inertial range depending on the global imbalance in the system is an interesting area for future research. Further investigation with better quality particle measurements or high-resolution numerical simulations is necessary to conclusively address this matter.

\subsection{Sensitivity of SDDA measurements to instrumental noise}\label{subsec:discuss_Instrumental}

In recent years, concerns have been raised regarding the extent to which our ability to estimate statistical quantities that depend on the sub-dominant Els\"asser mode is hampered by uncertainty in velocity field measurements due to instrument characteristics \citep{Podesta_2009, Chen_2012ApJ, PODESTA_2009_SDDA, 2012_Gogoberidze, bowen2021nonlinear, sioulas2024higherorder, Ervin_2024}. 

For example, the velocity data from the Wind mission are quantized before transmission back to Earth, a process that involves rounding real, rational numbers to the nearest integer. A close inspection of the velocity field time series shows that almost every data point has a time-discontinuity. While these discontinuities are small—corresponding to the bit size of the data—they induce an artificial $\propto f^{-2}$ spectrum. This high-frequency quantization noise leads to a decoupling of velocity and magnetic field fluctuations near the Nyquist frequency, identified as the primary reason for the observed decrease in $\sigma_c$ at high frequencies \citep{Podesta_2010}. Other sources of uncertainty, including aliasing and Poisson noise, are further discussed by \citet{2013_Chen_residual}.

Since good-quality, high-frequency velocity field measurements are not available at the moment, we have tried to quantify the uncertainty in our measurements to assess the degree of confidence we can place in our observations and to attribute them to physical mechanisms underlying the turbulent cascade.

In Section~\ref{subsec:instr-noise-modeling:theory}, we presented a simplified theoretical analysis along with an effort to model the effects of noise on SDDA measurements. Our analysis indicates that despite the fact that other quantities, e.g., $P_{\boldsymbol{v}}$, $\sigma_c$, or even $P_{z_{i}}$, may still be measurable for the largest portion of the inertial range, the effects of noise on alignment angle measurements can be dramatic even at very low frequencies. The strong dependence on the level of Els\"asser imbalance can render such measurements impossible even at scales as large as $5-10 \times 10^{3} d_i$. Therefore, while we discuss other alternative mechanisms of physical origin that could result in misalignment of the fields in subsequent sections, we believe, based on the analysis presented in Section~\ref{subsec:instr-noise:modeling:2}, that the increasing misalignment of the fields usually observed at inertial scales is of instrumental origin.

\subsection{Effects of compressibility in SDDA}\label{subsec:discuss_compress}

\par As the cascade progresses toward smaller spatial scales, the topology and characteristics of turbulent fluctuations can change dramatically. Various types of coherent structures, both incompressible (such as current sheets and Alfv\'en vortices) and compressible (including magnetic holes and solitons), have been observed in different space-plasma environments \citep{2006_alexandrova, 2006_Rees, 2016_perrone, Vasko_CS_21, vinogradov2023embedded}. In many cases, compressive fluctuations have been shown to significantly impact the dynamics of the turbulent solar wind \citep{Klein_2012, Howes_2012, Verscharen_2017, Bowen_2018_PDI, chandran_2018, 2019ApJ_Shoda}.

To bridge the gap between in-situ observations, theoretical predictions, and numerical evidence, we sought to quantify the effects of compressibility on SDDA. This involved a distinct examination of the dynamics of compressible and incompressible fluctuations to gain insights into the persistent observation of increasing misalignment at the inertial scale of solar wind turbulence.

As anticipated, our results indicated that strongly compressible fluctuations generally exhibit very weak to negligible signatures of increasing alignment, as shown in Figure~\ref{fig:compress_sdda}. These fluctuations can, to some extent, affect the average scaling of SDDA when mixed with Alfv\'enic fluctuations. However, compressible fluctuations are usually associated with weak field gradients and have considerably lower amplitudes than incompressible fluctuations, Figure ~\ref{fig:various_quants}. Therefore, their impact on the average scaling behavior is minimal, especially when amplitude-weighted definitions of the alignment angle are considered (see Equation~\ref{fin_ch_eq:8}).

In fact, though not shown here, the average behavior (i.e., without segregating compressible from incompressible fluctuations) of the alignment angles almost perfectly overlaps with the black (i.e., incompressible) lines in Figure~\ref{fig:compress_sdda}, indicating that the effects of compressibility on SDDA are negligible and can thus not explain the trend of misalignment at inertial scales.

\subsection{ Alignment at the Outer Scale: Assessing Consistency with Homogeneous, Incompressible MHD Phenomenologies}\label{subsec:outer_scale_alignment}

Our findings confirm a well-established result: increasing signatures of SDDA are observable for the majority of the intervals considered, irrespective of associated plasma parameters, at the outer scale. However, before concluding that the observed behavior aligns with the phenomenological models of homogeneous MHD turbulence discussed above, another aspect, which has not been extensively explored in the literature, warrants further discussion.

In homogeneous MHD, the turbulent cascade advances through uncorrelated collisions between counterpropagating wave packets. However, in the stratified solar wind, dynamics become more intricate due to linear couplings of outgoing waves with large-scale inhomogeneities, leading to non-WKB (Wentzel-Kramers-Brillouin) reflections. In this scenario, the Els\"asser fields can be decomposed into primary and secondary components \citep{velli_turbulent_1989, Velli_1990, Hollweg_Isenberg_2007}. The primary component, known as the ``classical'' component $z^{i}_{c}$, travels at the characteristic phase speed $V_{sw} - V_a$. The secondary, or ``anomalous''
component $z^{i}_{a}$, maintains the same phase function and thus propagating properties as the forcing, $z^{o}$. Consequently, $z^{i}_{a}$ maintains a strong correlation with $z^{o}$, $\boldsymbol{z}^{o} \propto - \boldsymbol{z}^{i}_{a}$. Numerical demonstrations of this anomalous coherence effect in inhomogeneous MHD turbulence have been explored by \cite{Verdini_2009ApJ} and \cite{Meyrand_2023}.

The nonlinear interactions between $z^{o}$ and $z^{i}_{c}$ are uncorrelated and transient, limited to the duration of their encounters. Conversely, in the frame of the outgoing wave, $z^{i}_{a}$ appears stationary, and the shearing between $z^{o}$ and $z^{i}_{a}$ remains coherent over time. It is intuitive to anticipate that as the imbalance in the fluxes of counterpropagating wavepackets increases, the efficiency of $z^{o}$ and $z^{i}_{c}$ decreases—it becomes more difficult for a $z^{i}_{c}$ fluctuation to locate and interact with the dominant $z^{o}$ fluctuations. Consequently, the influence of the anomalous coherence effect would intensify with increasing Els\"asser imbalance.

\begin{figure}[t]
     \centering
     \includegraphics[width=0.45\textwidth]{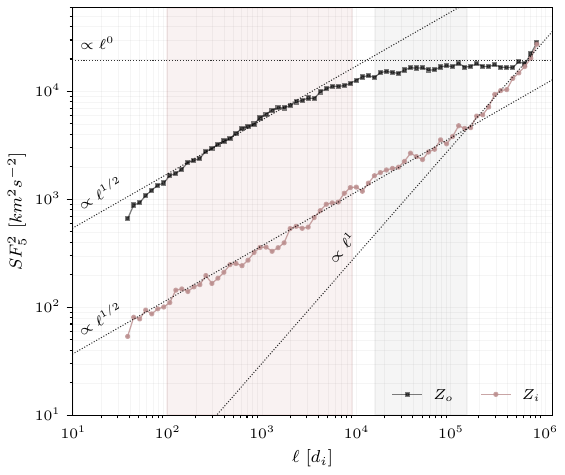}
     \caption{Averaged trace $SF_{5}^{2}$ of $\boldsymbol{z}_{o}$ (black) and $\boldsymbol{z}_{i}$ (red) for the fast, strongly Alfv\'enic ($\sigma_c>0.95$) solar wind. At inertial scales, indicated by the pink shading, both fields follow a $-3/2$ spectral scaling. Two outer scale regimes may be observed: at scales $\ell \in [2 \times 10^{4} - 2 \times 10^{5}] d_i$, spectral scalings for both fields, $P_{z_{o}} \propto k^{-1}$ and $P_{z_{i}} \propto k^{-3/2}$, are consistent with models based on ``anomalous coherence'' effects \citep{velli_turbulent_1989, Perez_Chandran_2013, Meyrand_2023}. At even larger scales, a range with $P_{z_{o}} \propto k^{-1}$ and $P_{z_{i}} \propto k^{-2}$ scalings is observed, consistent with the model by \citet{chandran_2018} based on PDI.}
     \label{fig:outer_scale_imbalanced intervals}
\end{figure}

The anomalous coherence of wave packets in the expanding solar wind enhances nonlinear interactions compared to the homogeneous scenario, thereby altering the phenomenology of the energy cascade. This results in $P_{z_{o}} \propto k^{-1}$ and $P_{z_{i}} \propto k^{-3/2}$ outer range scalings for inwardly and outwardly propagating modes, respectively \citep{velli_turbulent_1989, Perez_Chandran_2013, meyrand2023reflectiondriven}. By selecting the most strongly imbalanced ($\sigma_c(\ell^{\ast})>0.95$) fast wind intervals—having similar levels of RMS—and calculating an average second-order moment, we demonstrate that this is indeed the case, as shown in Figure~\ref{fig:outer_scale_imbalanced intervals}. Therefore, the coherent nature of the interactions could potentially affect the scale-dependence of the alignment angle. For example, the fact  that the anomalous fluctuations are highly aligned, and that the fraction of $z^{i}_{c}$ becomes larger as the cascade proceeds to  smaller scales, could provide an alternative explanation for the weakening or even reversal of SDDA.

This discussion underscores the need for a comprehensive theoretical framework to understand the nature of SDDA in the solar wind, taking into account the relative contributions of $z^{i}_{a}$ and $z^{i}_{c}$ to the shearing of $z^{o}$ across various scales. 


\subsection{Instabilities}\label{subsec:discuss_instabilities}

An important question arising from our study concerns the regime change observed in the polarization alignment, which tends to plateau, at best, as inertial scales are approached. An additional explanation for the observed increasing misalignment at smaller spatial scales involves the idea that dynamically aligned structures of a particular amplitude become unstable to disruption by tearing instabilities and the onset of magnetic reconnection \citep{Furth_tearing}. Particularly, when the maximum growth rate of the \cite{Coppi_1976} mode, $\gamma_{t}$, becomes comparable to the non-linear cascade time $\tau_{nl}$, $\gamma_{t} \tau_{nl} \gtrsim 1$, the current sheets can no longer remain stable \citep{Pucci_Velli_2014, Uzdensky_Loureiro_2016}. The disruption of the current sheets, at a certain scale $\lambda_{D}$, interrupts dynamic alignment, accelerating the turbulent cascade and leading to a pronounced steepening of the power spectrum \citep{2017_Mallet_tearing, 2017_Loureiro}. In the tearing-dominated regime, the alignment angle is expected to increase with decreasing scale, $\theta^{ub} \sim \lambda^{-4/5}$ \citep{Mallet_2017b, Boldyrev_Loureiro_2017, Comisso_2018ApJ}.

Using three-dimensional gyrokinetic simulations, \cite{2022_cerri} explored the dynamics of collisions between Alfv\'en waves (AW), revealing distinct behaviors in SDDA depending on whether the wave packets are colliding or well separated. When colliding, alignment demonstrates scale-dependent behavior likely induced by the packets shearing each other, whereas when well separated, a reconnection-mediated cascade dominates, leading to misalignment. This differentiation, as depicted in Figure 4 of \cite{2022_cerri}, is crucial; without it, the average curve lacks specific scaling and remains nearly flat. This averaging effect could significantly impact in-situ measurements, especially when alignment is measured over extended periods, exacerbating as the measurements move towards smaller scales where turbulence evolution timescales become shorter.

In many astrophysical scenarios, MHD turbulence is typically driven by localized sources (e.g., shear, instabilities), leading to non-balanced states, i.e., non-negligible cross helicity. Indeed, localized regions of imbalance have been prominently observed even in globally imbalanced simulations of MHD turbulence \citep{Matthaeus_2008_patches_imbalance, Perez_2009}. Patches of positive and negative cross-helicity are also evident in globally balanced solar wind streams \citep{Chen-anisotropic_2011, wicks_alignmene_2013}. Considering these observations and based on their simulations \cite{2022_cerri} propose that aligning thin, long-lived current sheets are generated by the turbulent cascade, then misaligning through tearing, in a patchy fashion in space and time, rather than stepwise in k-space. Taking these observations into account, we can conclude that while the tearing mediated regime is highly unlikely to occur at scales as large as $10^{3}-10^{4} d_i$, the impact of tearing of the current sheets on the alignment angle at inertial scales warrants further investigation.

In addition, other types of ideal MHD instabilities, such as the Kelvin-Helmholtz instability \citep{Malagoli_1996_KH}, could potentially manifest in the solar wind. For example, it has been demonstrated that despite the highly Alfv\'enic nature of the fluctuations at large scales, the signatures of Alfv\'enicity (namely, $\sigma_c, ~\sigma_r, ~\theta^{ub}$) diminish at considerably larger scales than those observed at 1 AU \citep{Podesta_2010}.

\cite{Parashar_2020} interpret these observations as indicative of substantial energy found in velocity shears, which disrupt an initial spectrum of high cross helicity by injecting equal amounts of the two Els\"asser energies \citep{Roberts_1987, Goldstein_1989, roberts_velocity_1992} within the inner heliosphere \citep{Ruffolo2020}. Since both $\sigma_c$ and $\sigma_r$ are interdependent with the alignment angles, such effects could potentially influence the SDDA scalings. While it is not possible to rule out such effects, especially given recent in-situ observations \citep{Paouris_2024}, it's essential to note that, as demonstrated in Section~\ref{subsec:discuss_Instrumental}, high-frequency instrumental noise in velocity field measurements can contaminate the power spectrum of the ingoing Els\"asser field even at lower frequencies. This contamination becomes progressively more significant as the Els\"ass\"er imbalance increases, see~\ref{fig:simulate_noise_level}b. Therefore, another plausible explanation for such observations may be contamination by high-frequency velocity field noise. Additionally, as pointed out in \citep{Schekochihin_2022}, since the vortex stretching terms for the different Els\"asser fields have opposite signs \citep{Zhdankin_2016}, there will generally be more ``current sheets'' than ``shear layers''. In such sheets, at least in the homogeneous case, the Kelvin–Helmholtz instability will be suppressed \citep{Chandrasekhar_1961}.

\subsection{Parametric Decay Instability \& ``Anomalous coherence'': Two Possible Outer Scale Regimes?}\label{subsec:discuss_PDI}

The coupling of large amplitude outwardly propagating waves to slow magnetosonic waves through the parametric decay instability \citep[PDI;][]{Galeev_1963, Goldstein_1978, 1997_prunareti_PDI, Reville_2018, Malara_22_PDI} results in two daughter waves: an antisunward propagating slow wave and a sunward propagating Alfv\'en wave of frequency slightly lower than the mother wave, could be another mechanism resulting in the misalignment of the fields. In the limit of low $\beta$, PDI is known to have a growth rate given by $\gamma/\omega_0 \sim \delta B/B_0 \beta^{-1/4}$ \citep{Galeev_1963_PDI, Goldstein_1989}, and thus its role could become important at the outer scale.

Considering the limit of strongly imbalanced, weak, compressible turbulence, \cite{chandran_2018} has demonstrated that PDI can lead to an inverse cascade of Alfv\'en wave quanta, resulting in outer-range scalings of $P_{z_{o}} \propto k^{-1}$ and $P_{z_{i}} \propto k^{-2}$. For this mechanism to become dominant over either the inward-outward interactions or the anomalous shearing discussed in Section~\ref{subsec:outer_scale_alignment}, the condition $\omega^{\pm}_{nl, ~PI} > \omega^{\pm}_{nl, \perp}$ must hold, where $\omega_{nl}$ represents the inverse of the respective dynamical timescales. Given the dependence of $\omega^{\pm}_{nl, ~PI}$ on $\delta B / B_0$, it is possible that two outer scale regimes may be realizable in the solar wind.

At large scales, $\lambda \gtrsim 10^{5} d_i$, where the weak-turbulence regime allows PDI effects to become dominant, increasing misalignment of the fields and thus lower $\sigma_c$ is observed towards larger scales, see e.g., Figure~\ref{fig:SDDA_sig_c}. It is noteworthy that magnetic compressibility monotonically increases with scale, see e.g., Figure~\ref{fig:compress_sdda}, providing further evidence supporting the idea that PDI might be dominant over this scale-range.

When considering the averaged trace $SF_{5}^{2}$ of $\boldsymbol{z}_{o}$ (black) and $\boldsymbol{z}_{i}$ (red) for the fast, strongly Alfv\'enic ($\sigma_c > 0.95$) solar wind, two outer scale regimes may be observed. At scales $\ell \in [2 \times 10^{4} - 2 \times 10^{5}] d_i$, the spectral scalings for both fields are consistent with models based on ``anomalous coherence'' effects \citep{velli_turbulent_1989, Perez_Chandran_2013, Meyrand_2023}. At even larger scales, a range with scalings consistent with the \citet{chandran_2018} model  is observed. This indicates that both mechanisms may be active in the solar wind, each becoming dominant at different scales.

Recent works considering Parker Solar Probe data \citep[PSP;][]{fox_solar_2016} have provided evidence for the dynamic formation of the $P_{z_{o}} \propto k^{-1}$ range of the power spectrum \citep{Davis_2023, Sioulas_2023_anisotropic, 2023_Huang}. While the PDI has been suggested as the plausible formation mechanism, the scaling of $P_{z_{i}}$ has not been considered, leaving open questions for future work. Additionally, it was shown that for $R < 0.3$ au and $\lambda \gtrsim 10^{4} d_i$, which roughly corresponds to the scale at which the $P_{z_{o}} \propto k^{-1}$ scaling was observed, turbulence is weak with $\chi^{\pm} \ll 1$ \citep{sioulas2024higherorder}. However, as the Alfv\'enic wavepackets propagate through the heliosphere, the decreasing Alfv\'en speed results in the growth of normalized magnetic-field amplitudes to nonlinear magnitudes, $\delta B/B_0 \sim 1$ \citep{Squire_2020, chen_evolution_2020, Mallet_2021, Tenerani_2021}, indicating that the regime of strong turbulence, $\chi^{-} \gtrsim 1$, might be extending to progressively larger scales with increasing heliocentric distance. This could allow for nonlinear shearing mechanisms to dominate over PDI, thus modifying the scaling of $P_{z_{i}}$.

Investigating the weak-to-strong turbulence transition \citep{Galtier_2000, Verdini_2012ApJ, Meyrand_transition}, the degree of compressibility, along with the scaling of both $P_{z_{i}}$ and $P_{z_{o}}$ in a range of different imbalance regimes, could provide a deeper understanding of the origin of the outer scale of solar-wind turbulence and the effects of the underlying physical mechanisms on SDDA. This will be the subject of future work.

\section{Summary \& Conclusions}\label{sec:Conclusions}

Using a large dataset of carefully selected homogeneous intervals from the WIND mission, we explored the impact of compressibility, intermittency, and imbalance on the statistical signatures of Scale-Dependent Dynamic Alignment (SDDA) in MHD turbulence within the solar wind \citep{boldyrev_2006, chandran_intermittency_2015}.

\vspace{1em}

Below, we summarize the key findings of our analysis. For clarity, we refer to the alignment angles of the Els\"asser variables as ($\Theta^{z}$) and the alignment between velocity and magnetic field fluctuations as ($\Theta^{ub}$).

\vspace{1em}
 (1)  SDDA in both $\Theta^{z}$ and $\Theta^{ub}$ is consistently evident at energy-containing scales, $\lambda \gtrsim 10^{4} d_i$; over the same scale range, $|\sigma_c|$ increases and $\sigma_r$ becomes more negative.

SDDA in both $\Theta^{z}$ and $\Theta^{ub}$ is consistently evident at energy-containing scales, $\lambda \gtrsim 10^{4} d_i$. The alignment increases as $|\sigma_c|$ increases and $\sigma_r$ becomes more negative.

\vspace{1em}
(2) $\Theta^{z}$ and $\Theta^{ub}$ exhibit an inverse correlation with the intensity of field gradients. This may stem from ``anomalous'' and/or ``counterpropagating'' wave packet interactions. Nevertheless, this observation indicates that the physical origin of alignment arises from the mutual shearing of the Els\"asser fields during imbalanced wave-packet interactions \citep{chandran_intermittency_2015}.

\vspace{1em}
(2) Compressible fluctuations do not exhibit any signs of SDDA. However, their effects on the average behavior of SDDA are negligible due to their relatively low amplitude and thus cannot explain the trend of misalignment at inertial scales.

\vspace{1em}
(3) Stringent thresholding on proxies for intermittency reveals SDDA signatures within a portion of the inertial range. Regardless of the approach we follow increasing misalignment is consistently  observed for scales $\lambda \leq 8 \times 10^{2} d_i$. Based on analytical arguments and modeling, we believe this misalignment is not a physical phenomenon but rather an artifact caused by high-frequency noise in the velocity field measurements.

\vspace{1em}
(4) The scaling of $\Theta^{z}$/$\Theta^{ub}$ becomes steeper/shallower with increasing global Alfv\'enic imbalance ($\sigma_r(\ell^{\ast}) \rightarrow -1$). However, only $\Theta^{ub}$ is correlated with global Els\'asser imbalance, becoming steeper with ($\sigma_c(\ell^{\ast}) \rightarrow 1$).

\vspace{1em}
(5) Signatures of increasing alignment in $\Theta^{ub}$ extend deeper into the inertial range of balanced intervals. This could be due to two factors: first, in globally imbalanced intervals, the fields are tightly aligned at large scales, depleting nonlinearities and halting the SDDA mechanism, causing alignment angles to saturate at a minimum value. Alternatively, instrumental noise may have a greater impact on alignment angle measurements as imbalance increases, leading to significant measurement errors.

\vspace{1em}
(6) Two outer scale regimes in the solar wind may be realizable: one dominated by parametric decay instability (PDI) \citep{chandran_2018} at larger scales where turbulence is weak ($\chi^{\pm} \ll 1$), and another by anomalous coherence effects \citep{velli_turbulent_1989} at intermediate scales where nonlinear Alfv\'enic interactions strengthen.

\vspace{2em}

While we do not disregard the need for theoretical advancements in existing turbulence models—especially to address the effects of imbalance and the expansion of the solar wind—it is more plausible that our results underscore the current limitations in precisely estimating alignment angles due to instrumental constraints, particularly in the context of highly Alfv\'enic, strongly  imbalanced turbulence. Unfortunately, the most intriguing cases are also the most difficult to investigate.

To further investigate the scaling of alignment angles, it is essential to consider alternative methodologies. For example, employing electric field measurements could facilitate the estimation of the perpendicular component of magnetic field fluctuations, given by $\delta \mathbf{V}_{\perp}(\mathbf{\ell}, t) = \delta \mathbf{E}_{\perp}(\mathbf{\ell}, t) \times \mathbf{B}_0/c$, where $\delta \mathbf{E}_{\perp}(\mathbf{\ell}, t)$ represents the perpendicular component of the electric field fluctuations and $c$ is the speed of light. This approach is planned for future research.

Moreover, 3D expanding box simulations of compressible MHD, which include both balanced and imbalanced turbulence, will offer deeper insights and facilitate more precise quantification of these effects on alignment measurements (Shi et al., in prep).

\vspace{12pt}

NS acknowledges insightful discussions with Dr. Lynn B. Wilson on the use of WIND data and with Dr. Michael Terres on MHD turbulence in the solar wind.

\vspace{0.1em}

This research was funded in part by the FIELDS experiment on the Parker Solar Probe spacecraft, designed and developed under NASA contract NNN06AA01C; the NASA Parker Solar Probe Observatory Scientist grant NNX15AF34G and the  HERMES DRIVE NASA Science Center grant No. 80NSSC20K0604. 

NS and MV were  supported by the International Space Science Institute (ISSI) in Bern, through ISSI International Team project. \# 550, Solar Sources and Evolution of the Alfv\'enic Slow Wind. T.A.B. acknowledges NASA Grant No. 80NSSC24K0272. BC was supported in part by NASA grant 80NSSC24K0171.

\software{Python \citep{van1995python}, SciPy \citep{2020SciPy-NMeth}, Pandas \citep{mckinney2010data},  Matplotlib \citep{Hunter2007Matplotlib}, PySpedas \citep{angelopoulos_space_2019}, MHDTurbPy \citep{MHDTurbPy_Sioulas}}

\clearpage
\appendix

\begin{figure*}[h]
     \centering
     \includegraphics[width=1\textwidth]{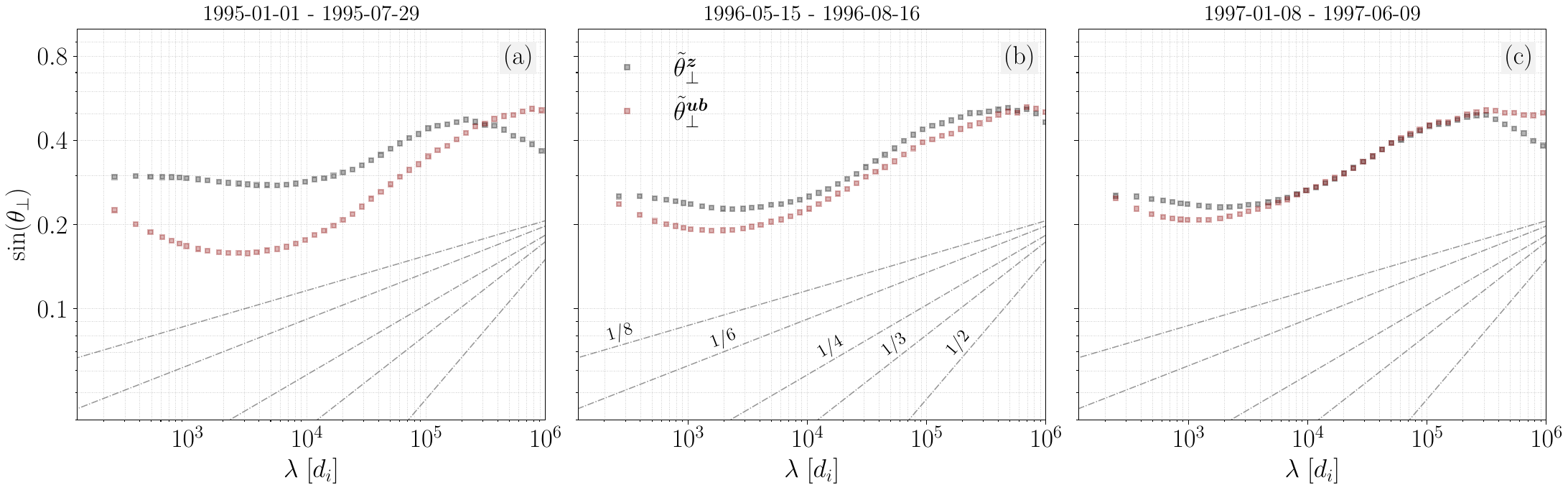}
     \caption{Alignment angle measurements as defined by Equations~\ref{fin_ch_eq:7} and ~\ref{fin_ch_eq:8} in top and bottom panels, respectively, across the three (out of four) intervals studied in \citep{PODESTA_2009_SDDA}.Black curves represent the alignment between Els\"sser fields, while red curves illustrate the alignment between velocity and magnetic field fluctuations. }
     \label{fig:2}
\end{figure*}

 \subsection{Revisiting \cite{PODESTA_2009_SDDA}}\label{appendix:1}

To verify the effectiveness of our modified alignment angle estimation method in comparison with \citetalias{Podesta_2009}, we revisited the intervals analyzed in their study. Their investigation primarily focused on the alignment between $\delta \boldsymbol{v}_{\perp}$ and $\delta \boldsymbol{b}_{\perp}$, employing Equation \ref{fin_ch_eq:8}. In addition they introduced a weighted average angle, defined in Equation~ 8 of \citetalias{Podesta_2009}. Our analysis indicates that the latter definintion results in scalings that are in most cases identical to Equation \ref{fin_ch_eq:8}.

Figure \ref{fig:2} showcases our findings for $\Tilde{\theta}_{\perp}^{ub(z)}$, with each column representing one of the three (out of four) intervals analyzed by \citetalias{Podesta_2009}. The Els\"asser fields are denoted by black lines, while the angles between magnetic and velocity fields are depicted in red. In all instances, the scaling obtained aligns with those reported in \citetalias{Podesta_2009}, with both the 2-pt and 5-pt increment methods yielding very similar results, as detailed in Section \ref{sec:data_analysis}. Thus, we can confidently proceed with our analysis utilizing the 5-pt increment method.

\subsection{Geometrical Constrains and Types of Alignment}\label{appendix:2}

To provide a clear illustration of the interdependence between $\sigma_c$, $\sigma_r$, $\theta^{ub}_{\perp}$, and $\theta^{z}_{\perp}$  we provide a graphical representation of the equations:

\begin{equation}\label{fin_ch_eq:geometry}
\cos(\theta^{z}_{\perp}) = \frac{\sigma_r}{\sqrt{1- \sigma_c^{2}}}, ~~ \cos(\theta^{ub}_{\perp}) = \frac{\sigma_c}{\sqrt{1- \sigma_r^{2}}},
\end{equation}

\begin{figure*}[h]
     \centering
     \includegraphics[width=0.75\textwidth]{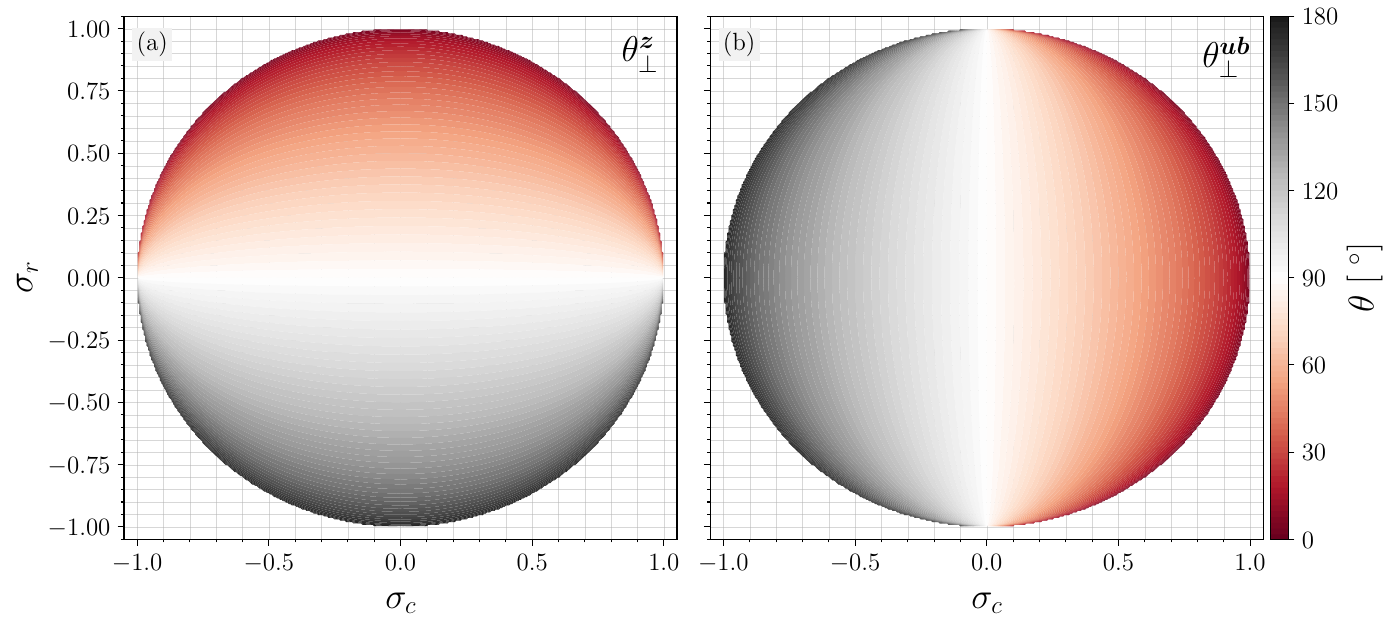}
     \caption{Graphical representation of Equations~\ref{fin_ch_eq:geometry} illustrated in panels (a) and (b), respectively.}
     \label{fig:sig_theta_relationship}
\end{figure*}

To generate random values for $\sigma_c$ and $\sigma_r$ within a unit circle on the Cartesian plane, constrained to the range of [-1, 1], we employed a random sampling approach. We independently selected values for $\sigma_c$ and $\sigma_r$ from a continuous uniform distribution spanning [-1, 1], while ensuring that the condition $\sigma_c^{2} + \sigma_r^{2} \leq 1$ was met to keep them within the unit circle. Subsequently, for each pair, we estimated the alignment angles according to Equation~~\ref{fin_ch_eq:geometry}. The results are illustrated in Figure \ref{fig:sig_theta_relationship}, colored by $\theta^{z}_{\perp}$ in panel (a), and $\theta^{ub}_{\perp}$ in panel (b).

\subsection{Modeling High Frequency Instrumental Noise: Perfectly aligned fields}\label{appendix:synthetic-timeseries-aligned}

We generate two sets of three-component Gaussian white noise time series for the magnetic field ($\mathbf{B}$) and the velocity field ($\mathbf{V}$), with the corresponding components being perfectly correlated ($\rho = 1$). The duration of the time series is set to 48 hours ($d = 48 \text{ hours}$), with a cadence of 3 seconds ($dt = 3 \text{ s}$), mimicking the typical duration and cadence of our identified intervals.  The base time series are transformed to the frequency domain using the Fast Fourier Transform (FFT). To shape the power-spectral density (PSD) of the time series, we apply filters in the frequency domain. For a given desired spectral exponent $\alpha$, the filter response $H(f)$ is defined as:

\begin{equation}\label{fin_ch_eq:spectral_shaping}
H(f) = |f|^{\alpha/2},
\end{equation}

where $f$ denotes the frequency.

 The spectral shaping filters are applied by multiplying the Fourier-transformed series by the filter response $H(f)$. The filtered series are then transformed back to the time domain using the inverse FFT. At this point, two time series with the desired correlation and spectral properties have been constructed.

Next, we need to properly normalize the two time series to model different levels of Els\"asser imbalance. This requires input from in-situ observations. It is well known that the RMS of the fluctuations in both velocity and magnetic field increases with $\sigma_c$ \citep{Borovsky_2019, Pi_2020, Sioulas_2023_anisotropic}. To quantify this dependency, we estimate the RMS value of the fluctuations in the different fields for each of our selected intervals. For any given field $\boldsymbol{\Phi}$, fluctuations are estimated as:

\begin{equation}
     \Delta \boldsymbol{\Phi} = \boldsymbol{\Phi} - \boldsymbol{\Phi}_{0},
     \label{fin_ch_eq:flucts}
\end{equation}

where $\boldsymbol{\Phi} = V_{a}, V, Z_{o}, Z_{i}$ and $\boldsymbol{\Phi}_{0} = \langle  \boldsymbol{\Phi} \rangle_D$ is a moving average with a window of duration $D = 8 \text{ hours}$. Subsequently, the RMS of each field is estimated using $d = 1 \text{ minute}$ moving averages of the fluctuations:

\begin{equation}
     \boldsymbol{\Phi}_{RMS} = \sqrt{ \langle \Delta \boldsymbol{\Phi}^{2} \rangle_d}.
\end{equation}

\begin{figure}[t]
     \centering
     \includegraphics[width=0.4\textwidth]{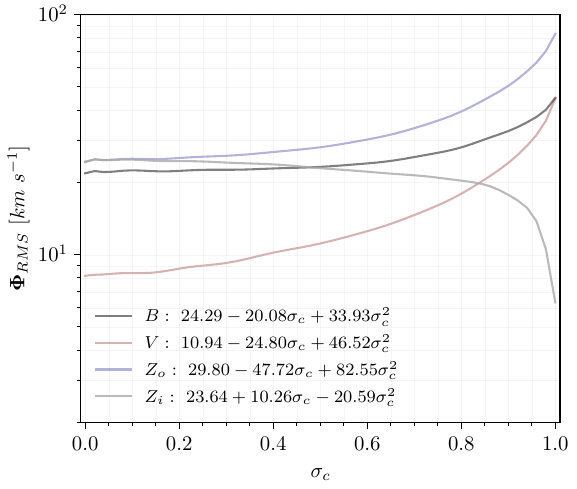}
     \caption{The empirical relationship for the RMS of the fluctuations is shown for (a) $\boldsymbol{B}$ (black line), (b) $\boldsymbol{V}$ (red line), (c) $\boldsymbol{z}_{o}$ (blue line), and (d) $\boldsymbol{z}_{i}$ (gray line), plotted against $\sigma_c$. Using the fluctuations estimated through Eq. \ref{fin_ch_eq:flucts}, 1-minute moving averages of $\sigma_c$ were also estimated, yielding a dataset of size $N \approx 7.5 \times 10^{7}$. The RMS quantities of the fields were plotted against $\sigma_c$, and an empirical fit was extracted by finding the best 12th-degree polynomial fit. The fits obtained for the different fields are plotted against $\sigma_c$, with the legends showing only the first three terms of the empirical fit.}
     \label{fig:all_fields_RMS}
\end{figure}

Using the fluctuations estimated through Eq. \ref{fin_ch_eq:flucts}, 1-minute moving averages of $\sigma_c$ are then also estimated. This process yielded a dataset of size $N \approx 7.5\times 10^{7}$ total measurements. $\boldsymbol{\Phi}_{RMS}$ are then plotted against $\sigma_c$, and an empirical fit is extracted by finding the best 12th-degree polynomial fit. The fit obtained for the different fields is plotted against $\sigma_c$ in Figure~\ref{fig:all_fields_RMS} with the legends showing only the first three terms.

Using the empirical relationship for the RMS of the fields as a function of $\sigma_c$, the time series are normalized such that the RMS of the synthetic data matches that of the real data at different levels of imbalance while retaining the original correlation and desired spectral scalings. To simplify our analysis, we assume—though this is strictly true only for the velocity field \citep{Chen_2012ApJ, Sioulas_2022_spectral_evolution, mcintyre2023properties}—that the inertial-range trace power-spectral scalings of the two fields are independent of $\sigma_c$ and scale as $a_{b} = a_{v} = -3/2$.

Small amplitude, high-frequency noise is then added to the $\boldsymbol{V}(t)$ series. We assume, though this is not strictly true \citep[see, e.g.,][]{Ervin_2024}, that the noise floor in the velocity field measurements is uniform across all intervals considered and independent of $\sigma_c$. Given that the RMS of the fluctuations peaks at $\sigma_c \approx 1$, we define the noise level as $\Delta = \alpha \boldsymbol{V}_{\text{RMS}}^{\ast}(f_{\text{Nyq}}) \mathcal{N}(0,1)$, where $\boldsymbol{V}_{\text{RMS}}^{\ast}(f_{\text{Nyq}})$ is the RMS value of the velocity field at the Nyquist frequency for $\sigma_c = 1$, and $\mathcal{N}(0,1)$ represents a normal distribution with mean, $\mu =  0$ and standard deviation $\sigma =1$. Different values of $\alpha$ have been considered. In the following, we examine the case that was found to provide the most realistic results, with $\alpha = 0.1$, corresponding to an error of $\Delta \approx 1 \, \text{km/s}$.

For each simulated interval, the Fourier trace power spectral density $F(f)$ was calculated, smoothed by averaging over a sliding window with a factor of 2, and then transformed into a wavenumber spectrum expressed in physical units $P(k d_i)$:

\begin{equation}
P(k d_i) = \frac{V_{sw}}{2 \pi d_i} F(f),
\end{equation}
where $k d_i = (2 \pi f d_i)/V_{sw}$ \citep{Sioulas_2022_spectral_evolution}. To perform this normalization, the ratio $V_{sw} / d_i$ was estimated for all intervals in our dataset, resulting in a mean value of $3.95 \pm 1.27 s^{-1}$. The mean value was used to normalize the synthetic spectra to enable a more meaningful comparison with the in-situ derived power spectra and spectral scaling indices.

\subsection{Modeling High Frequency Instrumental Noise: Correlation coefficient informed by in-situ observations.}\label{appendix:synthetic-timeseries-non- aligned}

The previous approach, by assuming perfect alignment between magnetic and velocity field fluctuations, restricts our analysis and does not allow $\sigma_c$ at large scales to decrease below $\sigma_c \lesssim 0.45$, thus preventing a fully consistent and realistic simulation. This limitation arises from the imposed correlation, which is not always observed in the solar wind.

We therefore need to follow a more rigorous approach. To this end, we generate time series for the magnetic and velocity fields that meet three simultaneous conditions informed by our in-situ observations: (1) the fluctuations are correlated with a specified correlation coefficient, (2) their inertial-range spectral scaling, and (3) the RMS values of these time series align with those observed in-situ. Therefore, in addition to considering the RMS values of the fields, the inertial-range spectral-scalings, $\alpha_{\phi}$, where $\phi = \boldsymbol{B},\boldsymbol{V}, ~\boldsymbol{Z}_{i}, ~\boldsymbol{Z}_{o}$ were estimated by finding the best-fit linear gradient in log–log space over the range $kd_i \in [5 \times 10^{-4}, 10^{-2}]$. In addition, the correlation coefficient

\begin{figure}[t]
     \centering
     \includegraphics[width=0.8\textwidth]{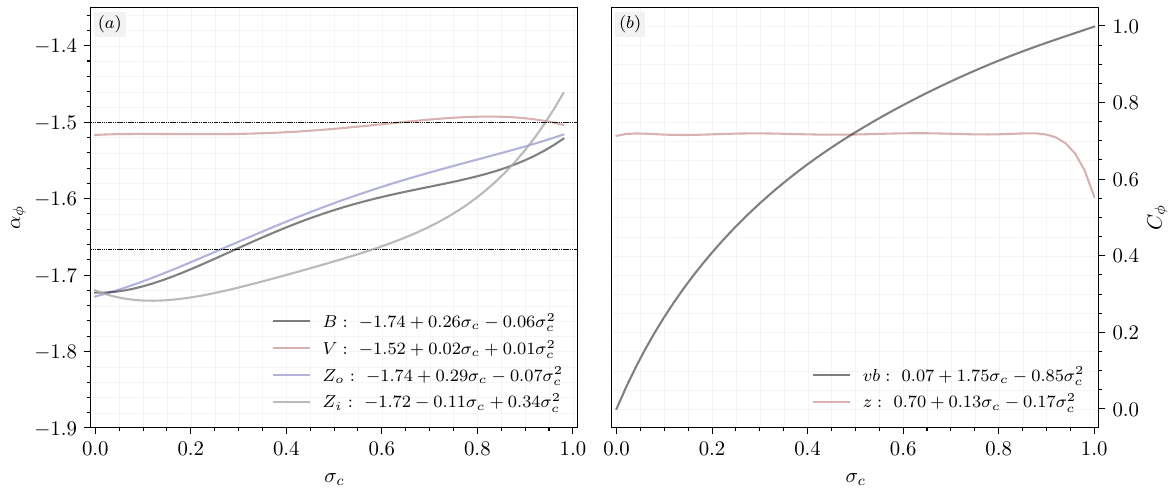}
     \caption{(a) The inertial-range power-spectral scalings for $\boldsymbol{B}$ (black line), $\boldsymbol{V}$ (red line), $\boldsymbol{z}_{o}$ (blue line), and $\boldsymbol{z}_{i}$ (gray line) plotted against $\sigma_c$, estimated by determining the best-fit linear gradient in log–log space over the range $kd_i \in [5 \times 10^{4}, 10^{-2}]$ for each selected interval in our dataset. (b) The correlation coefficient, $C_{\phi}$, as defined in Equation~\ref{fin_ch_eq:cor_coeff}, for the velocity/magnetic fields (black) and Els\"asser fields (red) plotted against $\sigma_c$. In contrast to the scaling index in panel (a), $C_{\phi}$ is estimated using the 1-minute moving average, similar to Figure~\ref{fig:all_fields_RMS}, from a dataset of size $N \approx 7.5 \times 10^{7}$.}
     \label{fig:all_fields_scalings_cor_coeff}
\end{figure}

\begin{equation}\label{fin_ch_eq:cor_coeff}
C_{\phi} = \frac{\langle \Delta \boldsymbol{\phi} \cdot \Delta \boldsymbol{\psi}\rangle}{\sqrt{ \langle \Delta \boldsymbol{\phi}^{2}\rangle \langle\Delta \boldsymbol{\psi}^{2} \rangle}},
\end{equation}
between either magnetic-velocity or Els\"asser field fluctuations, was estimated and an empirical relationship was obtained for these quantities as a function of $\sigma_c$. The obtained functional form for the spectral scalings and the correlation coefficients between the magnetic-velocity fields, $\rho_{ub}$, and Els\"asser fields, $\rho_{z}$, are illustrated in Figure~\ref{fig:all_fields_scalings_cor_coeff}.

While the third condition is straightforward to enforce and satisfying either of the first two conditions independently is also manageable, applying spectral shaping filters as described in Equation \ref{fin_ch_eq:spectral_shaping} to achieve the desired spectral scaling can disrupt the specified correlation coefficient, and vice-versa. Thus, imposing both conditions simultaneously in a rigorous manner is challenging.

To achieve this, we follow an iterative approach. Specifically, we generate time series for the magnetic field components $B_{i}$ with the desired spectral scaling of $a_B = -1.74 + 0.26 \cdot \sigma_c$. Each component $B_{i}$ of the magnetic field is paired with an initially uncorrelated component $V_{i}$ of the velocity field. The desired scaling index for the velocity field is $-1.5$, independent of $\sigma_c$. However, the iterative process to achieve the desired correlation coefficient modifies the imposed scaling. 

We have empirically found that if the initial scaling of the generated velocity time series is given by $a_v = -1.51 + 0.18 \cdot \sigma_c$, we can obtain both the desired scaling and correlation coefficient simultaneously through iterative adjustment. This process aims to match the desired correlation between the increments of $B_{i}$ and $V_{i}$. The input velocity field component is given by

\begin{equation}
V_{i} = \rho \cdot B_{i} + \sqrt{1 - \rho^2} \cdot \eta_{i},
\end{equation}
where $\rho$ is the desired correlation coefficient, and $\eta_{i}$ is the initially generated velocity time series with the imposed scaling. This provides an initial guess for $V_{i}$.

 For each iteration, the current correlation coefficient $r$ between $B_{i}$ and $V_{i}$ increments over specified intervals is computed. If the absolute difference between the current and desired correlation coefficients is less than the specified tolerance, the iteration stops. Otherwise, an adjustment is applied to $V_{i}$ based on the correlation error, defined as

\begin{equation}
\delta = \gamma \times (\rho - r) \times B_{i},
\end{equation}
where $\gamma$ is the learning rate and $r$ is the current correlation coefficient. This process is repeated until the desired correlation coefficient is achieved.

Once the desired correlation is achieved, we normalize the RMS values of the magnetic and velocity fields to ensure they meet the specified conditions. 

By carefully balancing the spectral properties and correlation through this iterative process, we successfully produced time series for the magnetic and velocity fields that simultaneously satisfied the specified conditions and adhered to the desired spectral scalings and correlation coefficients.

\begin{figure*}[t]
     \centering
     \includegraphics[width=1\textwidth]{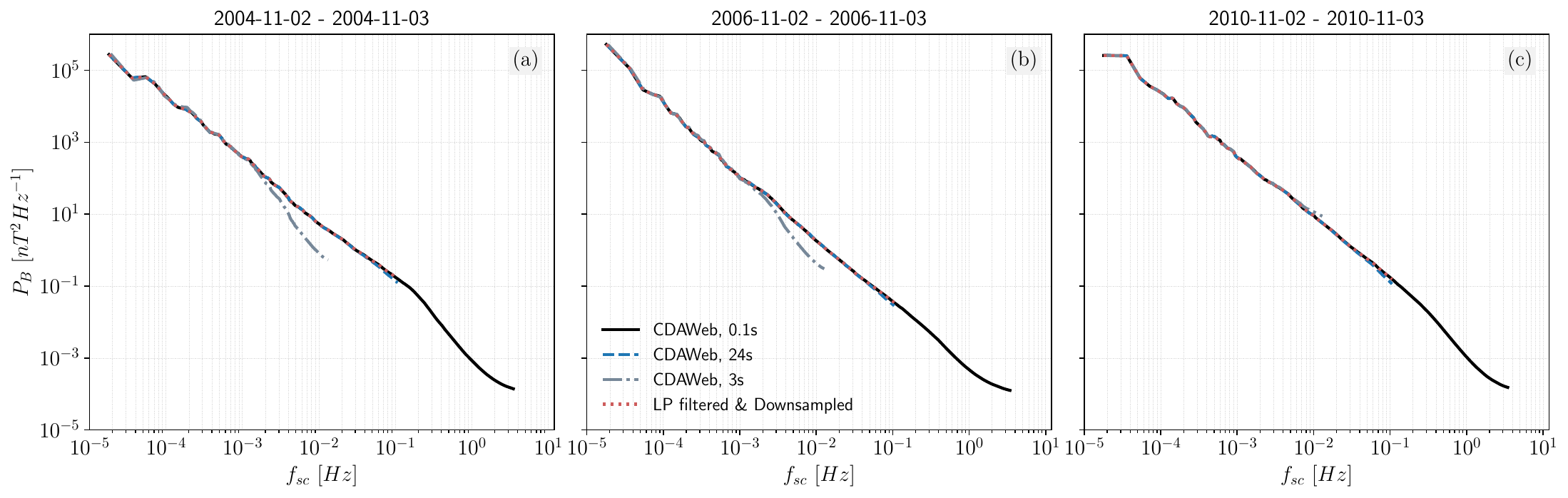}
     \caption{Trace power spectrum of the magnetic field time series for three randomly sampled 1-day intervals, analyzed using CDAWeb data at varying cadences. Displayed are the full resolution ($\delta \tau = 0.1s$, black line), downsampled data ($\delta \tau = 3s$, blue line; $\delta \tau = 24s$, cyan line), and full resolution data downsampled to the velocity field  cadence after processed with a 10th order \citep{BUTTERWORTH} low-pass filter (red line).
}
     \label{fig:downsampled_data}
\end{figure*}

\subsection{A problem with downsampled WIND data provided in CDAWeb}\label{appendix:CDWEB}

In digital signal processing, downsampling is commonly used to reduce the size of data sets, decrease computational demands , and align timeseries sampled at different cadence. However, to maintain the integrity of the signal, it is crucial to apply low-pass filtering before downsampling. According to the Nyquist-Shannon sampling theorem, to reconstruct a signal without loss of information, the sampling frequency, $f_s$, must be at least twice the highest frequency component, $f_{max}$, in the signal:

\begin{equation}
f_s \geq 2f_{max}
\end{equation}

Failing to adhere to this principle results in aliasing, where frequency components higher than $\frac{f_s}{2}$—the Nyquist frequency—appear as lower, incorrect frequencies in the sampled signal. Therefore, not applying a low pass filter before downsampling would result in an artificial flattening of the power-spectrum. Prior to downsampling, it is necessary to apply a low-pass filter that eliminates frequencies above the new Nyquist frequency. The cutoff frequency of this filter, $f_c$, should ideally be at or below $f_{s_{new}}/2$, where $f_{s_{new}}$ is the new, reduced sampling rate:

\begin{equation}
f_c \leq \frac{f_{s_{new}}}{2}
\end{equation}

The design of the low-pass filter, including its cutoff frequency and the steepness of its roll-off, significantly impacts the resulting signal. A filter with a steep roll-off efficiently attenuates frequencies near $f_c$, minimizing aliasing but possibly altering the natural characteristics of the signal. Conversely, a filter with a gentle roll-off may inadequately suppress high frequencies, resulting in a flatter than expected power spectral density (PSD). A well-designed filter ensures the PSD is shaped appropriately:

\begin{equation}
PSD(f) \approx \left\{
\begin{array}{ll}
S(f) & \text{if } f < f_c \\
0 & \text{if } f > f_c
\end{array}
\right.
\end{equation}

where $S(f)$ represents the original power spectral density of the signal. For example, if a signal sampled at 1 kHz is to be downsampled to 250 Hz, the original Nyquist frequency is 500 Hz, and the new Nyquist frequency post-downsampling is 125 Hz. Thus, a suitable low-pass filter should have a cutoff frequency at or below 125 Hz to prevent aliasing. Proper implementation of these filtering and downsampling steps preserves the fidelity of the signal, thereby ensuring the accuracy of any subsequent analysis.

While analyzing WIND data, we encountered issues with the downsampled magnetic field datasets available on CDWeb. Figure \ref{fig:downsampled_data} compares the trace power-spectral densities: the properly downsampled signal using a Butterworth low-pass filter of order $N=10$ is shown against the full-resolution dataset ($\delta \tau = 0.1s$, black line), the downsampled $\delta \tau = 3s$ dataset (blue line), and the downsampled $\delta \tau = 24s$ dataset (cyan line). As illustrated, both downsampled signals deviate from the original and appropriately downsampled power spectra. Although the deviations are less critical for the $\delta \tau = 3s$ dataset, they are considerably pronounced for the $\delta \tau = 24s$ dataset, potentially leading to erroneous interpretations of the turbulent cascade across the frequency range of $10^{-3}-10^{-2}$ Hz. We therefore recommend caution when utilizing these datasets and advise adopting a manual downsampling methodology if aligning the cadence of the magnetic and velocity field time series is required.

\bibliography{sample631}
\bibliographystyle{aasjournal}

\end{CJK*}
\end{document}